\documentclass[sn-basic]{sn-jnl}

\usepackage{graphicx}%
\usepackage{multirow}%
\usepackage{amsmath,amssymb,amsfonts}%
\usepackage{amsthm}%
\usepackage[title]{appendix}%
\usepackage{xcolor}%
\usepackage{textcomp}%
\usepackage{manyfoot}%
\usepackage{booktabs}%
\usepackage{algorithm2e}
\usepackage{listings}%
\usepackage{booktabs} 
\usepackage{comment}
\usepackage{multicol}
\usepackage{longtable}
\usepackage{tabularx}
\usepackage{subcaption}
\usepackage{lmodern}

\usepackage[margin=1in]{geometry}

\usepackage{dcolumn} 

\begin{document}

\title[Article Title]{Annealed variational mixtures for disease subtyping and biomarker discovery}


\author[1,2]{\fnm{Emma} \sur{Pr\'{e}vot}}\email{emma.prevot@exeter.ox.ac.uk}

\author[1,3]{\fnm{Rory} \sur{Toogood}}\email{rory.toogood@btinternet.com}

\author[1,4]{\fnm{Filippo} \sur{Pagani}}\email{filippo.pagani@warwick.ac.uk}

\author[1]{\fnm{Paul D. W.} \sur{Kirk}}\email{paul.kirk@mrc-bsu.cam.ac.uk}

\affil[1]{\orgdiv{MRC Biostatistics Unit}, \orgname{University of Cambridge}, \orgaddress{\street{East Forvie Building, Robinson Way}, \city{Cambridge}, \postcode{CB2 0SR}, \country{UK}}}

\affil[2]{\orgdiv{Department of Statistics}, \orgname{University of Oxford}, \orgaddress{\street{24-29 St Giles'}, \city{Oxford}, \postcode{OX1 3LB}, \country{UK}}}

\affil[3]{\orgdiv{Department of Mathematical Sciences}, \orgname{Durham University}, \orgaddress{\street{Stockton Road}, \city{Durham}, \postcode{DH1 3LE}, \country{UK}}}

\affil[4]{\orgdiv{Department of Statistics}, \orgname{University of Warwick}, \orgaddress{\city{Coventry}, \postcode{CV4 7AL}, \country{UK}}}

\abstract{Cluster analyses of high-dimensional data are often hampered by the presence of large numbers of variables that do not provide relevant information, as well as the perennial issue of choosing an appropriate number of clusters. These challenges are frequently encountered when analysing omics datasets, such as in molecular precision medicine, where a key goal is to identify disease subtypes and the biomarkers that define them. Here we introduce an annealed variational Bayes algorithm for fitting high-dimensional mixture models while performing variable selection. Our algorithm is scalable and computationally efficient, and we provide an open source Python implementation, {\em VBVarSel}. 
In a range of simulated and real biomedeical examples, we show that VBVarSel outperforms the current state of the art, and demonstrate its use for cancer subtyping and biomarker discovery. 
}




\maketitle

Precision medicine has the potential to revolutionise the treatment and prevention of diseases by tailoring healthcare strategies and therapies to the specific characteristics of particular groups of individuals \citep{Wang2023}. A significant factor driving advances in precision medicine, particularly in oncology \citep{bigdataPrecisionmed}, is the wealth of high-dimensional omics datasets. These datasets are frequently analysed using clustering algorithms to identify disease subtypes, followed by a characterisation of the clusters in terms of the biomarkers that define them, with the ultimate goal of deepening our understanding of diseases and improving patient outcomes \citep{golub1999molecular, cancer2013tcga}. However, this process faces substantial challenges due to the high dimensionality and heterogeneity of the data \citep{witten2010framework,review, kirk2023bayesian}.

In many applications, it has been common to include all available variables in the modeling process, based on the assumption that using all the information will enhance the performance of a clustering algorithm \citep{law2004simultaneous}. However, in practice, this approach can be computationally expensive, and may negatively impact the stratification process by including irrelevant or ``noisy" variables \citep{hassie19, BOUVEYRON201452, Gnanadesikan95}. To address this, variable selection techniques can be employed to simplify result interpretation and enhance the quality of data classification \citep{review}.

Bayesian approaches, such as Bayesian mixture models, have proven effective in facilitating both stratification and feature selection in high-dimensional unsupervised settings, where there are no available labels to guide selection and subtyping \citep{review}. Many existing algorithms rely on Markov Chain Monte Carlo (MCMC) methods \citep{bensmailMCMCbook} for inference, as these can quantify uncertainty in the allocation of observations to clusters, as well as the number of clusters. However, MCMC is computationally demanding and tends to scale poorly in high-dimensional settings, making its application to real biomedical datasets often slow and impractical, or requiring dimensionality reduction. 
In contrast, Variational Inference (VI) typically provides a more scalable and efficient alternative for inference \citep{Blei_2017}, although does not by itself address the issue of the presence of (large numbers of) irrelevant variables.

There are several existing algorithms for variable selection in clustering \citep[for reviews, see][]{review2, review3, review}, most of which are either heuristic, rely on maximum likelihood estimation via the expectation maximisation (EM) algorithm, or use MCMC. For example, $K$-means sparse clustering \citep[available in the sparcl R package][]{witten2013sparcl} alternates between fixing variable weights to optimise clustering and adjusting them based on the between-cluster sum of squares (BCSS). While computationally efficient in high-dimensional data, it requires the number of clusters either to be known in advance or to be optimised by considering a range of values for $K$, which comes at a computational cost.  The Sequential Updating and Greedy Search (SUGS) algorithm \citep{LianmingSUGS,sugs1}, which avoids computationally costly MCMC methods for Bayesian clustering using a sequential, greedy algorithm, was recently adapted to the variable selection setting \citep[SUGSVarSel;][]{CrookGattoKirk+2019}, but this greedy approach may sometimes lead to suboptimal solutions. 
The PReMiuM R package \citep{liverani2014premium} uses MCMC to perform inference for Dirichlet process mixtures that may permit variable selection, but is computationally intensive. 
Other approaches, like VarSelLCM \citep{VarSelLCM} and VSCC \citep{andrews2014variableVSCC}, optimise information criteria such as the Bayesian Information Criterion (BIC) or integrated complete-data likelihood (MICL). However, these methods often require navigating a large and combinatorially complex state space, which can limit their scalability. A VI-based approach for clustering and variable selection (`feature saliency') was provided by \citet{constantinopoulos2006bayesian}. However, the authors do not report run-times or provide a generally useable implementation, and the method's performance was primarily assessed on image classification tasks with small sample sizes, which may not generalize well to high-dimensional biomedical datasets.  The existing methods highlight the need for a more flexible and computationally efficient solution, capable of handling the complexity and scale of modern biomedical datasets.

A frequently overlooked challenge in both MCMC and VI is the issue of local optima, where it becomes exceedingly difficult, if not impossible, to escape local optima in multi-modal landscapes. As a result, performance is highly dependent on the number of local optima in the objective function, the initialisation of the algorithm, and the quality of the assumptions on the prior probability distributions. A general approach to addressing the local optima problem and improving inference is through simulated annealing \citep{ROSE1990589, ueda1998deterministic, Katahira2008DeterministicAV}. Rooted in principles from statistical mechanics, annealing introduces a temperature parameter into the objective function, which is gradually adjusted according to a time-dependent schedule. This smooths the objective function, helping to prevent the optimisation process from getting trapped in shallow local optima.

In the context of mixture models, we find that annealing can be particularly beneficial due to the inherent multi-modality of these models. However, to the best of our knowledge, no existing literature has provided a comprehensive annealing framework specifically tailored to our problem context. \citet{bayesianTadesse} mention the use of parallel tempering \citep{earl2005parallel} for Bayesian mixture models, but do not provide mathematical or empirical details. In contrast, \cite{helene} offer a detailed annealed VI framework, focused on variable selection and show that it provides more robust and stable inference, but only consider regression tasks with numerous predictors and multiple outcomes.

In this work, we present a scalable and computationally efficient algorithm for simultaneous clustering and variable selection, utilizing Variational Inference. We demonstrate the scalability of our method, making it suitable for extremely large and high-dimensional datasets, while maintaining accuracy, reliability, and good performance. Additionally, we incorporate annealing into the algorithm to improve inference when handling multi-modal posterior distributions, which are frequently encountered in the settings of this work.

\section*{Methods}\label{sec1}
\subsection*{Statistical model}\label{sec11}

We approach the task of subtyping via unsupervised model-based clustering. The model we adopt is a finite mixture of probability distributions, where each distribution, or component, characterises a distinct cluster \citep{baseMMFraley, lau2007, modelBasedClus}. Unlike conventional techniques (e.g.\ k-means \citep{macqueen1967kmeans}, hierarchical clustering \citep{ward1963hierarchical}), Bayesian mixtures offer a robust statistical framework for a probabilistic interpretation of cluster allocations \citep{modelBasedClus}, which is particularly relevant in biomedical applications. Importantly, as we follow the unsupervised learning approach, our model will discover hidden structures (clusters) within unlabeled data.

\subsubsection*{Finite mixture model}

Using a notation similar to \citet{bishopML}, let $X= \left \{ {\bf x}_n \right \}_{n=1}^N$ be the data matrix, where ${\bf x}_n$ is a $J$-dimensional vector of random variables, $J$ being the number of features. A finite mixture model with $K$-components (clusters) is defined as
\begin{equation} \label{eq:basicClustModel}
    p(X| \Phi,\pi) = \prod_{n=1}^N \prod_{k=1}^K \pi_k f_X({\bf x}_n|\Phi_{k}) \, ,
\end{equation}
where $\pi_k$ is the mixture weight corresponding to component $k$. That is to say, $\pi_k$ is the probability that observation ${\bf x}_n$ was generated from $f_X({\bf x}_n|\Phi_{k})$, the individual likelihood for the mixture component $k$, which depends on the set of parameters $\Phi_{k}$. Once the data has been generated from the model above, the goal is to infer which observations (${\bf x}_i$) were generated from which cluster component.

A persistent challenge in clustering methods, including mixture models, is selecting the maximum number of clusters $K$ appropriately. 
Selecting a $K$ that is lower than the true number of clusters in the data leads to model misspecification. Selecting $K$ to be much larger than the true number of clusters results in wasted computational effort. 
We adopt an approach similar to \citep{rousseau2011asymptotic}, using an overfitted mixtures with a large but finite $K$, where the ``extra'' components are shrunk towards zero during the inference process.

\subsubsection*{Feature selection}

Clustering becomes more challenging when the data are high-dimensional and heterogeneous \citep{kirk2023bayesian}. While using all the available features can theoretically improve clustering performance \citep{law2004simultaneous}, it is often suboptimal from the perspective of computational efficiency. Some features may be unrelated to the clustering structure and add noise to the process \citep{MIAO2016919, featureSelSurv}. Additionally, using all the features contributes to well-known problems such as over-parametrisation, and the curse of dimensionality \citep{DynamicProgramming, Gnanadesikan95, hassie19, BOUVEYRON201452}. Hence, employing variable selection techniques can reduce the computational burden, improve model fit, and simplify the interpretation of results \citep{review}. 

For our purpose, algorithms can be divided into two broad classes: filter methods, which involve variable selection as a separate pre-processing or post-processing step, and wrapper methods, where variable selection is integrated within the learning process \citep{review, featureSelSurv}. 
We concentrate on wrapper methods, specifically Bayesian variants, as they can simultaneously perform variable selection and clustering, an approach that has several benefits \citep{kirk2023bayesian}.
In this framework, we model variable selection via a set of latent variables that determine whether the covariate contributes to the clustering structure or not. Simultaneously, we infer cluster assignments from the posterior based on the subset of active covariates.

Starting from the mixture model in Equation \eqref{eq:basicClustModel}, we assume that conditional on the clustering allocation, the covariates are independent from each other. This allows us to factorise the functional form $f({\bf x}_n | \Phi_k)$ as follows:
\begin{align}
f({\bf x}_n | \Phi_k) &= \prod_{j = 1}^J f_j(x_{nj}|\Phi_{kj}) \, ,
\end{align} 
with a separate $f_j(x_{nj}|\Phi_{kj})$ for each covariate.
We then introduce a latent binary variable $\gamma_j \in \left \{ 0,1\right \}$, indicating whether feature $j$ should be used to infer the clustering structure ($\gamma_j=1$) or not ($\gamma_j=0$). 
The likelihood becomes
\begin{align}
f({\bf x}_n | \Phi_k, \gamma) &= \prod_{j = 1}^J f_j(x_{nj}|\Phi_{kj})^{\gamma_j}f_j(x_{nj}|\Phi_{0j})^{1 - \gamma_j} \, ,
\end{align} 
where $\Phi_{0j}$ denotes parameter estimates obtained under the null assumption that the $j^{th}$ feature is not relevant to the clustering structure\footnote{The parameters $\Phi_{0j}$ are easily computable via Maximum Likelihood, and in order to increase computational efficiency, we can sample from the posterior conditional on the values of the precomputed $\{ \Phi_{0j} \}_{j=1}^J$.}. Note that the clustering structure only depends on relevant covariates, and is independent of irrelevant ones \citep{review}.

\subsection*{Variational Inference}

Our approach relies on Variational Inference (VI), which we also refer to as Variational Bayes (VB), an optimisation-based method that minimises a function of the Kullback-Leibler (KL) divergence between the true posterior $p(\theta | X)$, and a flexible and tractable approximation $q(\theta)$. The KL divergence \citep{kullback1951information} is a measure of the difference between two distributions, and can be defined as
\begin{align}
KL(q||p) &= -\int q(\theta) \ln \left(\frac{p(\theta|X)}{q(\theta)} \right) d\theta \, ,
\end{align} 
where $KL(q||p) \ge 0$, with equality holding if and only if $q(\theta) = p(\theta | X)$. 
However, minimizing the KL divergence directly is a difficult task. Instead, researchers typically aim to find the $q(\theta)$ that maximizes the Evidence Lower Bound (ELBO),
\begin{align}
\mathcal{L}(q) &= \int q(\theta) \ln \left(\frac{p(X, \theta)}{q(\theta)} \right) d\theta \, ,
\end{align} 
which is a quantity intrinsically related to the KL divergence in the following way:
\begin{align}
\ln p(X) = \mathcal{L}(q) + KL(q||p) \, .
\end{align}

While the choice of $q(\theta)$ is arbitrary, we follow a framework that originated in physics and is known as mean-field theory \citep{parisi1979toward}. In order to simplify the optimisation problem, we define $q(\theta)$ as
\begin{align}
q(\theta) = \prod_{i = 1}^M q_i(\theta_i) \, .
\end{align}
The optimisation process then simplifies to the iterative refinement of each factor $q_i(\theta_i)$ of $q(\theta)$, based on the current estimates of the other factors, following the equation
\begin{align}
\label{qUpdate}
q_l^\ast(\theta_l) = \frac{\exp\left( \mathbb{E}_{i\ne l}[\ln p(X, \theta)]\right)}{\int \exp\left( \mathbb{E}_{i\ne l}[\ln p(X, \theta)]\right)d\theta_l} \, .
\end{align} 
As the problem is convex with respect to each factor $q_l(\theta_l)$, the ELBO increases with every iteration of Equation \eqref{qUpdate} and is guaranteed to converge to a local optimum \citep{boyd2004convex}.

\subsection*{Annealing}

One major challenge in VI is escaping poor local optima, which can prevent the algorithm from finding the global optimum \citep{ ROSE1990589, bayesianTadesse}. This is particularly relevant in clustering and biomarker identification, where multiple plausible clustering structures may exist \citep{kirk2023bayesian}. In order to address this, we use a technique called annealing, which is based on principles of statistical mechanics and maximum entropy, and can help navigate intricate posterior landscapes \citep{ROSE1990589, ueda1998deterministic, Katahira2008DeterministicAV}. We introduce a temperature parameter $T$ into the ELBO that allows us to gradually transition from the prior to the (approximate) posterior while we explore the state space. The annealed version of the ELBO can be written as:
\begin{align}
\label{eq:annealing-update}
    \mathcal{L}(q) = \int q(\theta) \ln p(X, \theta) d\theta - T\int q(\theta) \ln q(\theta) d\theta
\end{align}

When $T=1$, Equation \eqref{eq:annealing-update} reduces to the standard ELBO, which encourages exploitation in the exploration/exploitation dichotomy. When $T>1$ the term corresponding to the prior distribution (which for obvious reasons is known as the ``entropy term") is given more weight, while in relative terms, the log of the joint distribution loses weight. As a result, increasing the temperature causes the variational distribution $q(\theta)$ to align more closely with the prior. This has the effect of flattening the variational objective function, which helps to prevent the optimization from becoming trapped in shallow local optima \citep{mandt2016variational}.

The update equation for $q_l^\ast(\theta_l)$ with annealing can be written as follows:
\begin{align}
q_l^\ast(\theta_l) = \frac{\exp\left( \frac{1}{T}\mathbb{E}_{i\ne l}[\ln p(X, \theta)]\right)}{\int \exp\left( \frac{1}{T}\mathbb{E}_{i\ne l}[\ln p(X, \theta)]\right)d\theta_l} \, .
\end{align}

\subsection*{The VBVarSel algorithm}

In this paper we present the algorithm VBVarSel, which uses annealed VI and variable selection to improve on the performance of the state-of-the-art methods on challenging clustering problems.
Algorithm \ref{alg:VBVarSel} below shows the pseudocode implementation of our algorithm, and we provide full mathematical details in Section \ref{app-sec-methods} of the Supplementary Material. 

The algorithm begins by initialising the variational parameters and calculating initial parameter estimates via Maximum Likelihood. A crucial component of VBVarSel is the annealing temperature schedule, which dictates how the temperature $T$ is initialized, set, and varied throughout the inference process. Since there is no consensus in the literature on the optimal type of schedule, we determine our approach empirically. We implement fixed schedules, as well as both geometric and harmonic temperature schedules to balance exploration and exploitation during optimization.  For the geometric schedule we follow \citet{helene} and \citet{kirkpatrick1983optimization}. The temperature at each iteration \( i \) is defined as:
\[
T_i = T_0 \alpha^i,
\]
where \( T_0 \) is the initial temperature and \( \alpha \) is the cooling rate calculated as:
\[
\alpha = \left( \frac{1}{T_0} \right)^{1 / (i_a - 1)}.
\]
This ensures that the temperature gradually decreases to \( T = 1 \) after a specified number of annealed iterations \( i_a \). We also implement a harmonic schedule for a slower, more gradual decline in temperature:
\[
T_i = \frac{T_0}{1 + \alpha i},
\]
with the cooling rate:
\[
\alpha = \frac{T_0 - 1}{i_a}.
\]
Both schedules facilitate a ``balancing act'' between \textit{exploration} and \textit{exploitation}. High temperatures in early iterations encourage exploration, allowing the algorithm to find various configurations and avoid getting trapped in shallow local optima. As iterations progress, decreasing the temperature shifts the focus toward exploitation, enabling the algorithm to refine and optimize the best solutions found so far.

Importantly, when using a \textit{fixed temperature} greater than one, as suggested by \citet{Katahira2008DeterministicAV} and \citet{mandt2016variational}, the inference targets an annealed approximate posterior throughout. In contrast, using either the geometric or harmonic schedule ultimately retrieves the same approximate posterior as non-annealed inference since the temperature gradually decreases to \( T = 1 \). This distinction is accounted for during empirical comparisons to assess the effectiveness of each schedule.

Throughout the iterative process, VBVarSel updates the variational parameters, evaluates cluster allocations \( Z = \{ z_n \}_{n=1}^N \) and variable selection indicators \( \gamma = \{ \gamma_j \}_{j=1}^J \), and computes the Evidence Lower Bound (ELBO) to monitor convergence. The algorithm continues until convergence criteria are met or a maximum number of iterations is reached. It is important to note that during our implementation, we addressed several potential numerical instabilities to prevent underflow and overflow, ensuring the robustness of our algorithm. Full mathematical details and parameter initialization procedures are provided in Section~\ref{app-sec-methods} of the Supplementary Material.

\RestyleAlgo{ruled}
\begin{algorithm}[hbt!]
\caption{The VBVarSel algorithm}\label{alg:VBVarSel}
\KwIn{Data $X = \left \{x_n\right \}_{n=1}^N $, maximum number of clusters $K$, temperature schedule,}
\hspace*{1.2cm} initial temperature $T_0$, maximum iterations $itr_{max}$, convergence threshold $\epsilon$
\BlankLine
\KwOut{Cluster allocations $Z = \left \{z_n\right \}_{n=1}^N $ }
\hspace*{1.5cm} Variable selection indicators $\gamma = \left \{\gamma_j\right \}_{j=1}^J $
\BlankLine

Initialise variational parameters;

Calculate parameter estimates for $\Phi_{0j}$ via Maximum Likelihood;
\BlankLine
$converged \gets False$;

$i \gets 0$;
\BlankLine
\While{$i < itr_{max}$}{
\eIf{T\_schedule is geometric \text{OR} harmonic}{
    $T \gets \text{eval\_temp\_schedule}(T_0, i, i_a)$
}{
    $T \gets T_0$ 
 }
 \BlankLine
Update variational parameters

Evaluate $Z$ and $\gamma$
\BlankLine
Compute ELBO according to Eq. \eqref{eq:elbo}

$improve \gets \text{ELBO}[i] - \text{ELBO}[i-1]$ 
\BlankLine
\If{$i>0$ \text{and} $0 < improve < \epsilon$}{
    $converged \gets \text{True}$
    
    \textbf{break}
}
\BlankLine
$i \gets i + 1$
}

\end{algorithm}

\subsubsection*{Performance evaluation}
When applying VBVarSel to new datasets for the first time, we experiment with various parameter initializations and temperature schedules to determine the optimal configuration. We assess each setup by monitoring the Evidence Lower Bound (ELBO) and select the one that maximizes it. After identifying the best parameter initialization for a specific experiment, we run VBVarSel for 10–20 repetitions using this chosen configuration. To mitigate the influence of stochastic elements inherent in the inference process, we randomize the ordering of data covariates in each repetition. This approach helps ensure that our results are robust and not dependent on a particular data arrangement or random seed.
Performance evaluation was twofold. First, a qualitative analysis was carried out through visualisation tools such as scatter plots and heatmaps. Then a quantitative assessment was done using the Adjusted Rand Index (ARI) \citep{RAND, Hubert1985ComparingP}, and an analysis of selected versus discarded covariates.  The Rand index (RI) is a measure of similarity between two data clusterings evaluated as the number of pairs of observations that are either in the same or different clusters in both partitions. The RI is \textit{adjusted} to account for the fact that some agreement can occur by chance using the formula:
\begin{align}
    \text{ARI} = \frac{\text{RI} - \mathbb{E}[\text{RI}]}{\text{max(RI)} - \mathbb{E}[\text{RI}]} \, .
\end{align}

The ARI's range from -1 to 1, where 1 indicates perfect agreement, 0 indicates random agreement, and $<0$ corresponds to assignments that are worse than random. For the quantitative evaluation of our experiments, we report median scores, together with upper and lower quartiles, evaluated on 10-20 repetitions.

\section*{Results}\label{sec2}

\subsection*{Simulation study}

We first assessed the performance of our algorithm in simulation studies where the ``ground truth" is known.
We replicated the simulation setup from \citep{CrookGattoKirk+2019}, which uses three $p$-dimensional Gaussian distributions with mixing weights 0.5, 0.3, and 0.2. These distributions have spherical covariance, and are centred at $(0, 0, ..., 0), (2, 2, ... , 2)$, and $(-2, -2, ... , -2)$ respectively.
The irrelevant variables are generated according to standard Gaussian distributions.
We generated either $n=100$ or $1000$ data points, and while we kept the total number of variables fixed to 200, we varied the percentage of relevant variables to be either 5\%, 10\%, 25\%, or 50\%.
We compared the performance of VBVarSel to i) hierarchical clustering and ii) K-means sparse clustering from \textit{sparcl} R package \citep{witten2010framework}\citep{witten2013sparcl}, iii) SUGSVarSel \citep{CrookGattoKirk+2019}, iv) VSCC \citep{andrews2014variableVSCC},  and v) VarSelLCM \citep{marbac2017variableVARSELLCM}. The results are shown in Table \ref{crookSim5} through Table \ref{crookSim50}, and we provide details of prior specifications for the different algorithms in Table \ref{prior init} of the Supplementary Materials. 
For each table we show the runtime (in seconds), the proportion of both relevant and irrelevant variables that were correctly identified, and the Adjusted Rand Index (ARI) \citep{RAND, Hubert1985ComparingP} between the inferred clustering and the ground-truth labels from the generated data. Each cell in the tables shows the median value, and the upper and lower quartiles calculated over 10 repetitions of the experiment. 
All algorithms were executed on the same High Performance Cluster to ensure fair runtime comparisons. However, it should be noted that VBVarSel is implemented in Python, whereas the other methods are R packages with optimised C++ code, potentially giving them a speed advantage.

\begin{table}[htp!]
\centering
\begin{tabular}{lccccc}
\toprule
Method & $n$ & Time (seconds) & Relevant & Irrelevant & ARI \\
\midrule
\textbf{VBVarSel} & 100 & 1.30 [1.19, 2.79] & 1 [1, 1] & 1 [1, 1] & 0.99 [0.98, 1] \\
Hier-clust & 100 & 5.27 [4.89, 5.66] & 1 [1, 1]  & 1 [1, 1]  & 0.68 [0.64, 0.71] \\
K-means & 100 & 7.45 [6.88, 7.95] & 1 [1, 1]  & 1 [1,1] & 0.61 [0.60, 0.61]\\
SUGSVarSel & 100 & 9.71 [9.98, 8.85] & 0 [0, 0] & 0.72 [0.70, 0.75] & 0 [0, 0.01] \\
VSCC & 100 & 5.64 [3.28, 6.69] & 1 [0.92, 1] & 0.94 [0.93, 0.95] & 1 [0.97, 1] \\
VarSelLCM & 100 & 321 [306, 331] & 1 [1, 1] & 1 [1, 1] & 1 [0.97, 1] \\
\hline 
\textbf{VBVarSel} & 1000 & 27.5 [27.4, 27.5] & 1 [1, 1] & 1 [1, 1] & 0.95 [0.90, 0.96] \\
Hier-clust & 1000 & 530 [499, 539] & 1 [1, 1] & 1 [1, 1] & 0.79 [0.69, 0.85]\\
K-means & 1000 & 93 [90, 98]  & 1 [1, 1] & 1 [1, 1] & 0.61 [0.61, 0.61] \\
SUGSVarSel & 1000 &  224 [198, 246] & 0 [0,0] & 0.97 [0.95, 0.98] & 0 [0,0.01]\\
VSCC & 1000 & 99.7 [98.5, 101] & 1 [1, 1] &  0 [0, 0] & 0 [0, 0] \\
VarSelLCM & 1000 & 49582 [46786, 50090] & 1 [1, 1] & 1 [1, 1] & 1 [1, 1] \\
\bottomrule
\end{tabular}
\caption{\label{crookSim5} \textbf{Simulation performance with 5\% relevant variables}. This table shows the results of applying VBVarSel on data simulated following the setup in \citep{CrookGattoKirk+2019}, where \textbf{5\%} of the variables are relevant. The cells show the median, lower and upper quartile of each variable, calculated over 10 repetitions.}
\end{table}

\begin{table}[htp!]
\centering
\begin{tabular}{lccccc}
\toprule
Method & $n$ & Time (seconds) & Relevant & Irrelevant & ARI \\
\midrule
\textbf{VBVarSel} & 100 & 3.23 [1.30, 3.23] & 1 [1, 1] & 1 [0.99, 1] & 1 [0.98, 1] \\
Hier-clust & 100 & 4.46 [4.19, 4.60] & 0.85 [0.80, 0.85] & 1 [1, 1] & 0.69 [0.68, 0.75]  \\
K-means & 100 & 7.51 [7.01, 8.11] & 1 [1, 1] & 1 [1, 1] & 0.62 [0.61, 0.63] \\
SUGSVarSel & 100 & 9.00 [8.91, 9.46]  &  0 [0,0] & 0.69 [0.68, 0.71] & 0.02 [0,0.04]\\
VSCC & 100 & 4.55 [3.53, 6.60] & 1 [1, 1] & 0.97 [0.96, 0.98] & 1 [1, 1] \\
VarSelLCM & 100 & 313 [298, 327] & 1 [1, 1] & 1 [1, 1] & 1 [1, 1] \\
\hline 
\textbf{VBVarSel} & 1000 & 27.6 [27.5, 28.3] & 1 [1, 1] & 1 [0.99, 1] & 0.92 [0.87, 0.99] \\
Hier-clust & 1000 & 450 [438, 480] & 0.85 [0.85, 0.85] & 1 [1, 1] & 0.74 [0.71, 0.88] \\
K-means & 1000 & 95 [90, 97]  &  1 [1, 1] & 1 [1, 1] & 0.61 [0.61, 0.61] \\
SUGSVarSel & 1000 & 211 [206, 249]  & 0 [0, 0] & 0.97 [0.96, 0.98] & 0 [0, 0] \\
VSCC & 1000 & 100 [99.6, 102] & 1 [1, 1] & 0 [0, 0] & 0 [0, 0] \\
VarSelLCM & 1000 & 52993 [52321, 53738] & 1 [1, 1] & 1 [1, 1] & 1 [1, 1] \\
\bottomrule
\end{tabular}
\caption{\label{crookSim10} \textbf{Simulation performance with 10\% relevant variables}. This table shows the results of applying VBVarSel on data simulated following the setup in \citep{CrookGattoKirk+2019}, where \textbf{10\%} of the variables are relevant. The cells show the median, lower and upper quartile of each variable, calculated over 10 repetitions.}
\end{table}

\begin{table}[htp!]
\centering
\begin{tabular}{lccccc}
\toprule
Method & $n$ & Time (seconds) & Relevant & Irrelevant & ARI \\
\midrule
\textbf{VBVarSel} & 100 & 3.22 [3.21, 3.23] & 1 [1, 1] & 1 [1, 1] & 1 [1, 1] \\
Hier-clust & 100 & 4.14 [3.86, 4.63] & 0.59 [0.54, 0.62] & 1 [1, 1] & 0.73 [0.66, 0.75] \\
K-means & 100 & 7.63 [7.36, 8.60] & 0.99 [0.98, 1] & 1 [1, 1] & 0.62 [0.61, 0.62] \\
SUGSVarSel & 100 & 7.81 [7.24, 7.94]  & 0 [0, 0] & 0.73 [0.70, 0.75] & 0.01 [0, 0.03]\\
VSCC & 100 & 2.05 [1.98, 2.20] & 1 [1, 1] & 0 [0, 0] & 1 [1, 1] \\
VarSelLCM & 100 & 296 [295, 305] & 1 [1, 1] & 1 [1, 1] & 1 [1, 1] \\
\hline 
\textbf{VBVarSel} & 1000 & 27.5 [27.5, 27.5] & 1 [1, 1] & 1 [0.99, 1] & 1 [1, 1] \\
Hier-clust & 1000 & 429 [401, 484] & 0.96 [0.95, 0.96] & 1 [1, 1] & 0.74 [0.69, 0.88] \\
K-means & 1000 & 94 [90, 99] & 1 [1, 1] & 1 [1, 1] & 0.61 [0.61, 0.61] \\
SUGSVarSel & 1000 & 162 [130, 192]  &1 [1, 1] & 1 [1, 1] & 1 [1, 1]\\
VSCC & 1000 & 103 [97.0, 107] & 1 [1, 1] & 0 [0, 0] & 0 [0, 0] \\
VarSelLCM & 1000 & 65652 [64364, 66513] & 1 [1, 1] &1 [1, 1]  & 1 [1, 1] \\
\bottomrule
\end{tabular}
\caption{\label{crookSim25} \textbf{Simulation performance with 25\% relevant variables}. This table shows the results of applying VBVarSel on data simulated following the setup in \citep{CrookGattoKirk+2019}, where \textbf{25\%} of the variables are relevant. The cells show the median, lower and upper quartile of each variable, calculated over 10 repetitions.}
\end{table}

\begin{table}[htp!]
\centering
\begin{tabular}{lccccc}
\toprule
Method & $n$ & Time (seconds) & Relevant & Irrelevant & ARI \\
\midrule
\textbf{VBVarSel} & 100 & 3.25 [3.24, 3.26] & 1 [1, 1] & 1 [1, 1] & 1 [1, 1] \\
Hier-clust & 100 & 4.48 [4.23, 4.67] & 0.34 [0.30, 0.35] & 1 [1, 1] & 0.72 [0.67, 0.85] \\
K-means & 100 & 7.84 [7.34, 8.35] & 0.46 [0.45, 0.48] & 1 [1, 1] & 0.61 [0.60, 0.62] \\
SUGSVarSel & 100 & 6.25 [6.18, 92.7]  & 0 [0, 0.49] & 0.75 [0.54, 0.79] & 0 [0, 0.61]\\
VSCC & 100 & 1.20 [1.03, 1.24] & 1 [1, 1] & 0 [0, 0] & 1 [1, 1] \\
VarSelLCM & 100 & 369 [353, 381] & 1 [1, 1] & 1 [1, 1] & 1 [1, 1] \\
\hline 
\textbf{VBVarSel} & 1000 & 27.8 [27.7, 27.9] & 1 [1, 1] & 1 [0.99, 1] & 1 [1, 1] \\
Hier-clust & 1000 & 499 [469, 523] & 0.44 [0.42, 0.45] & 1 [1, 1] & 0.71 [0.68, 0.77] \\
K-means & 1000 & 94 [93, 100] &  0.48 [0.48, 0.51] & 1 [1, 1] & 0.61 [0.61, 0.61] \\
SUGSVarSel & 1000 & 82.6 [77.6, 100]  &1 [1, 1] &1 [1, 1] & 1 [1, 1] \\
VSCC & 1000 & 173 [169, 181] & 1 [1, 1] & 0.82 [0.18, 0.95] & 0.95 [0.11, 1] \\
VarSelLCM & 1000 & 80599 [77222, 81929] & 1 [1, 1] & 1 [1, 1] & 1 [1, 1] \\
\bottomrule
\end{tabular}
\caption{\label{crookSim50} \textbf{Simulation performance with 50\% relevant variables}. This table shows the results of applying VBVarSel on data simulated following the setup in \citep{CrookGattoKirk+2019}, where \textbf{50\%} of the variables are relevant. The cells show the median, lower and upper quartile of each variable, calculated over 10 repetitions.}
\end{table}

In all simulations with $n$ = 1000, VBVarSel is consistently the fastest method, outperforming the second fastest method by a factor of at least 2.5. When $n$ = 100, VBVarSel continues to exhibit significant speed advantages, with only two cases where VSCC is marginally faster. However, despite its competitive runtime, VSCC performs poorly, often selecting all the variables without discrimination, resulting in clustering structures based on noise rather than signal.

In terms of clustering and feature extraction performance, VBVarSel and VarSelLCM are the only two methods that consistently achieve perfect results. For both moderately small ($n$ = 100) and larger ($n$ = 1000) simulated datasets, these methods can identify all relevant and irrelevant variables, and are able to recover the true clustering structure. However, VarSelLCM is at least at least 90 times slower when $n$ = 100, and at least 2000 times slower than VBVarSel when $n$ = 1000. 
This is due to the fact that when the true number of clusters is unknown, VarSelLCM needs to conduct an exhaustive search for a good estimate of this parameter, leading to high computational costs. In contrast, VBVarSel can automatically infer the number of clusters in the data, making it more efficient.

While K-means and Hierarchical Clustering (Hier-clust) have longer but manageable runtimes, they tend to be overly conservative in variable selection, often missing relevant variables, which negatively impacts their clustering accuracy. SUGSVarSel also has longer yet manageable runtimes, but performs poorly despite being evaluated in the simulation scenario proposed by its authors \citep{CrookGattoKirk+2019}.

Overall, VBVarSel stands out as the fastest and most scalable method, drastically reducing runtime despite the inherent computational limitations of its programming language. It achieves this while maintaining high, and often perfect, accuracy in both stratification and variable selection.

Furthermore, VBVarSel exhibited a remarkable level of robustness to model misspecification, as shown in Tables \ref{tab:misspecification 1} and \ref{tab:misspecification 2} in the Supplementary Materials. Tables \ref{annealing simCrook-corr-rand-all},  \ref{annealing simCrook-corr-fix}, \ref{annealing simCrook-corr-rand}, \ref{annealing simCrook}, and \ref{annealing simCrook-noise} in the Supplementary Materials instead demonstrate the benefits of annealing in overcoming common challenges faced when analysing real-world data, such as correlated data, sub-optimal parameter initialisation, and noise.

\subsection*{Breast cancer transcriptomic subtyping}

Breast cancer, the most common cancer among women worldwide \citep{NHSbreast}, exhibits diverse molecular traits and disease manifestations.  Research has identified various subtypes \citep{sorlie2003repeated, prat2010phenotypic,curtis2012genomic,  cancer2013tcga, duan2013metasignatures,  lock2013preprocess, akbani2014pan}, with a widely accepted classification dividing them into five categories based on three receptor proteins: estrogen receptor (ER), progesterone receptor (PR), and HER2 \citep{cancer2012comprehensiveBreast}. These subtypes are Luminal A, Luminal B, HER2-enriched, Basal-like, and Normal-like \citep{prat2010phenotypic, orrantia2022subtypes}. Stratification largely relies on gene expression profiles, notably the PAM50 gene set, which is crucial for both identifying subtypes and predicting risk \citep{sorlie2003repeated, parker2009supervisedBreast}.

For our analysis, we utilised breast cancer transcriptomic data from The Cancer Genome Atlas (TCGA), a comprehensive cancer genomics program that has molecularly characterised over 20,000 primary cancers, yielding extensive genomic, epigenomic, transcriptomic, and proteomic data \citep{cancer2013tcga}.

Our extracted dataset comprises transcriptomic information from 348 breast cancer tumor samples (rows), each with gene expression data for 17,814 genes (columns). After the removal of 441 genes with missing (NaN) entries, we refined the dataset to 17,373 genes across the same number of samples. Notably, our dataset included the 50 genes from the PAM50 set, a recognised gene set for breast cancer subtyping. Demographic and clinical characteristics, along with ‘ground-truth’ labels for cluster assignments into five breast cancer subtypes, were obtained from the Supplementary Table 1 of \citep{cancer2012comprehensiveBreast}. \\

We first applied VBVarSel the full 348 x 17373 TCGA transcriptomic breast cancer dataset \citep{cancer2013tcga}. It is worth noting that while most state of the art methods require various preprocessing steps due to scalability issues, our proposed algorithm converged in less than one hour.

\begin{figure}[htp] 
\centering
    \subfloat[On all covariates]{%
        \includegraphics[width=0.48\textwidth]{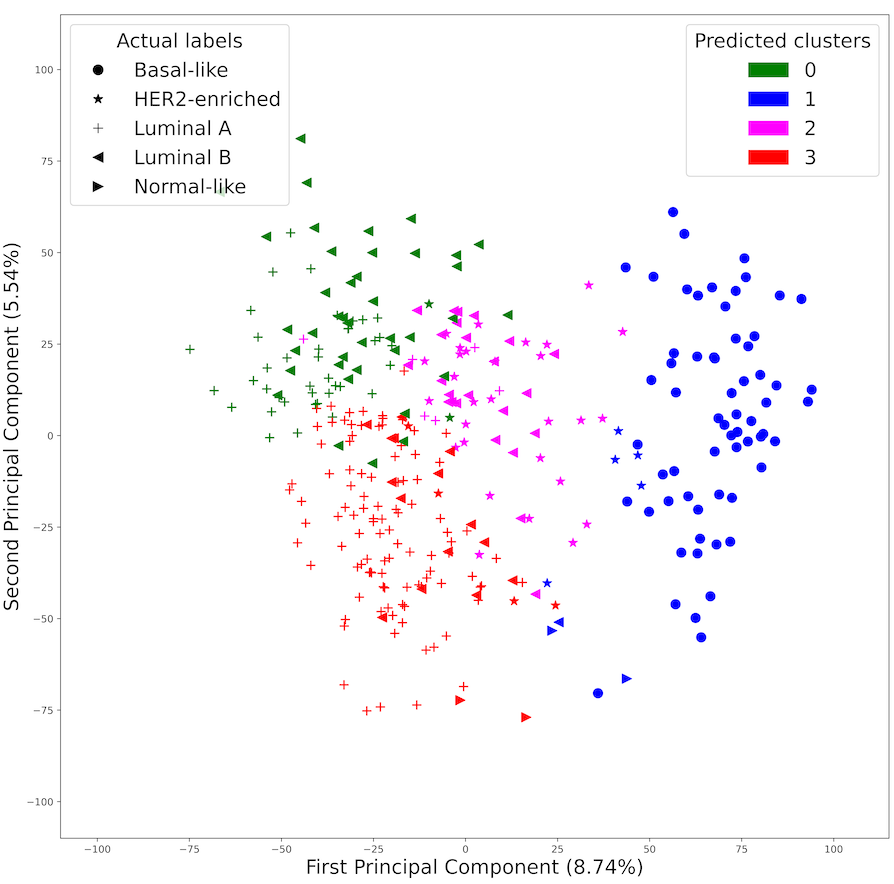}%
        }%
    \hfill %
    \subfloat[On only selected covariates]{%
        \includegraphics[width=0.48\textwidth]{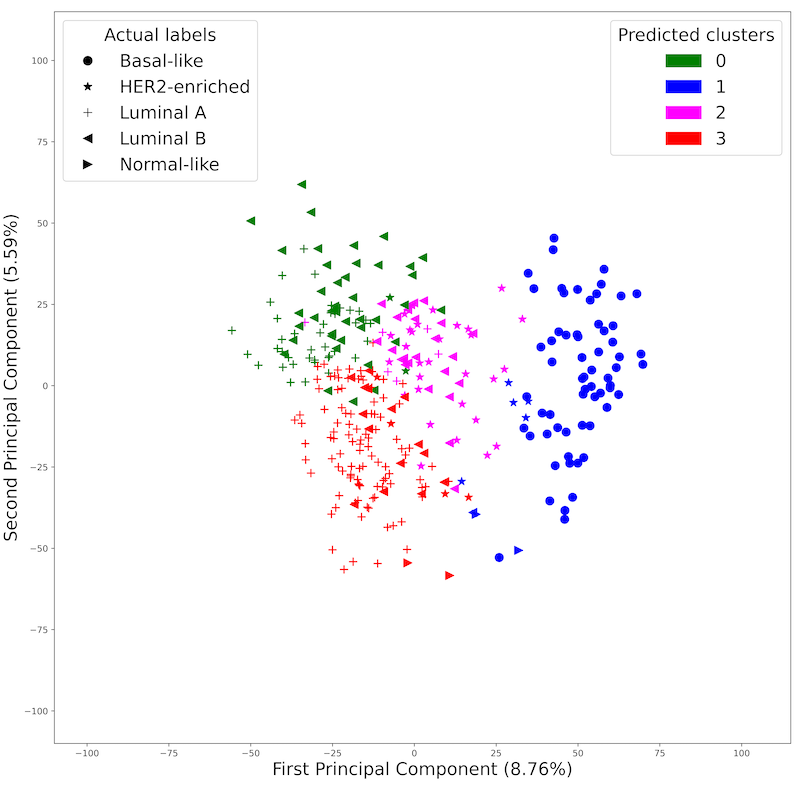}%
        }%

    \caption{\textbf{VBVarSel TCGA 4-cluster model}. This figure shows the scatter plots of VBVarSel 4-cluster model on the complete TCGA dataset, when PCA is applied to either all covariates (a), or only the selected ones (b).}
    \label{fig: scatter_cluster_alldataset_4clust all vs sel}
\end{figure}

In the final outcome, VBVarSel settled on a 4-5 clusters model where a median of 6723 covariates were selected, approximately 39\% of the full set. 
Among the PAM50 genes, which are regarded as informative for breast cancer subtyping, the median rate of selection was 75\%, which is similar to what was achieved with smaller dataset sizes, and significantly better than random (\textit{Fisher Test}, $p \ll 0.00001$). 
This stratification is medically sensible and is in line with the literature, as can be seen from Figure \ref{fig: scatter_cluster_alldataset_4clust all vs sel}.

\begin{figure}[htp] 
\centering
    \subfloat[On all covariates]{%
        \includegraphics[width=0.48\textwidth]{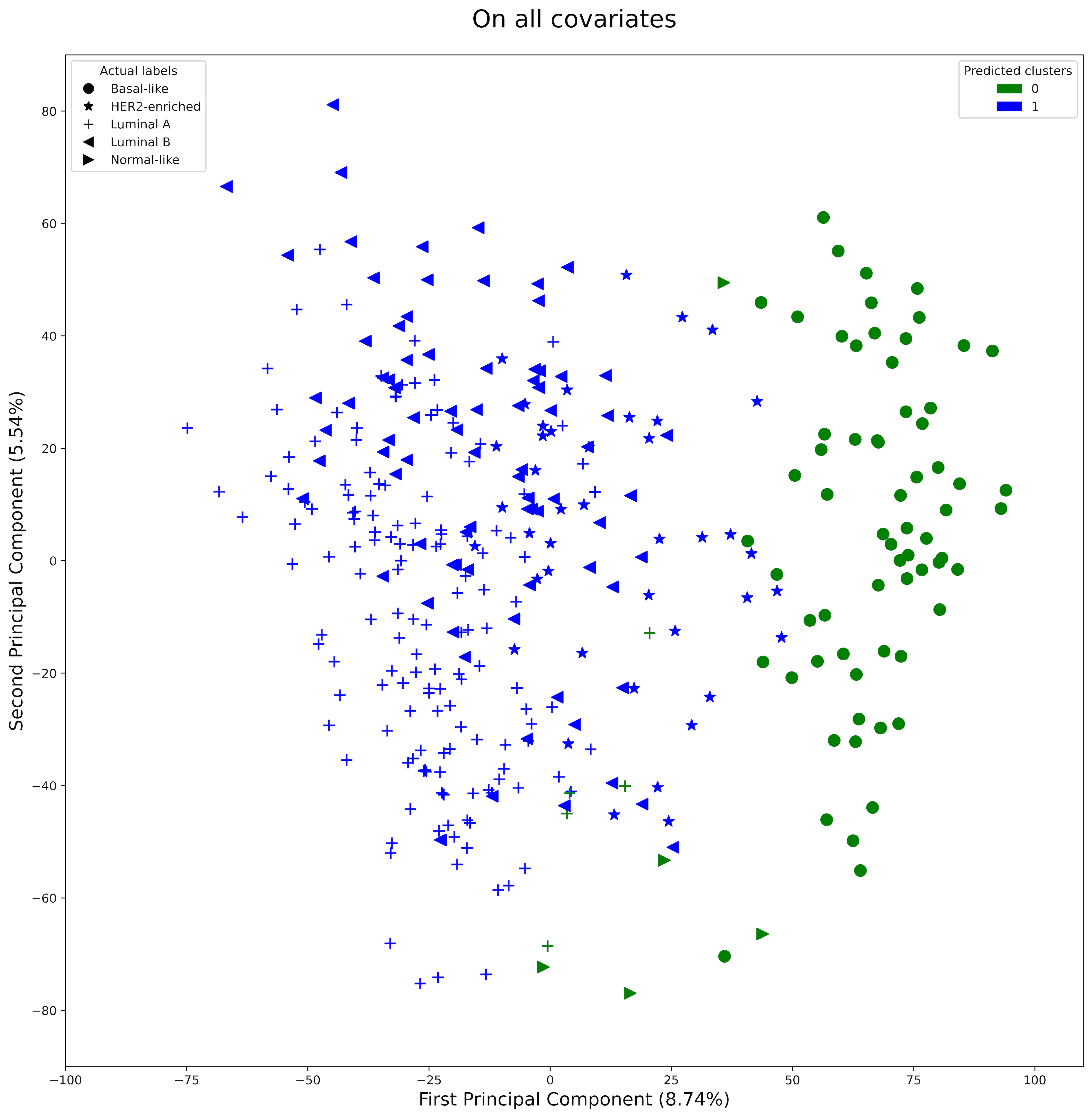}%
        }%
    \hfill %
    \subfloat[On only selected covariates]{%
        \includegraphics[width=0.48\textwidth]{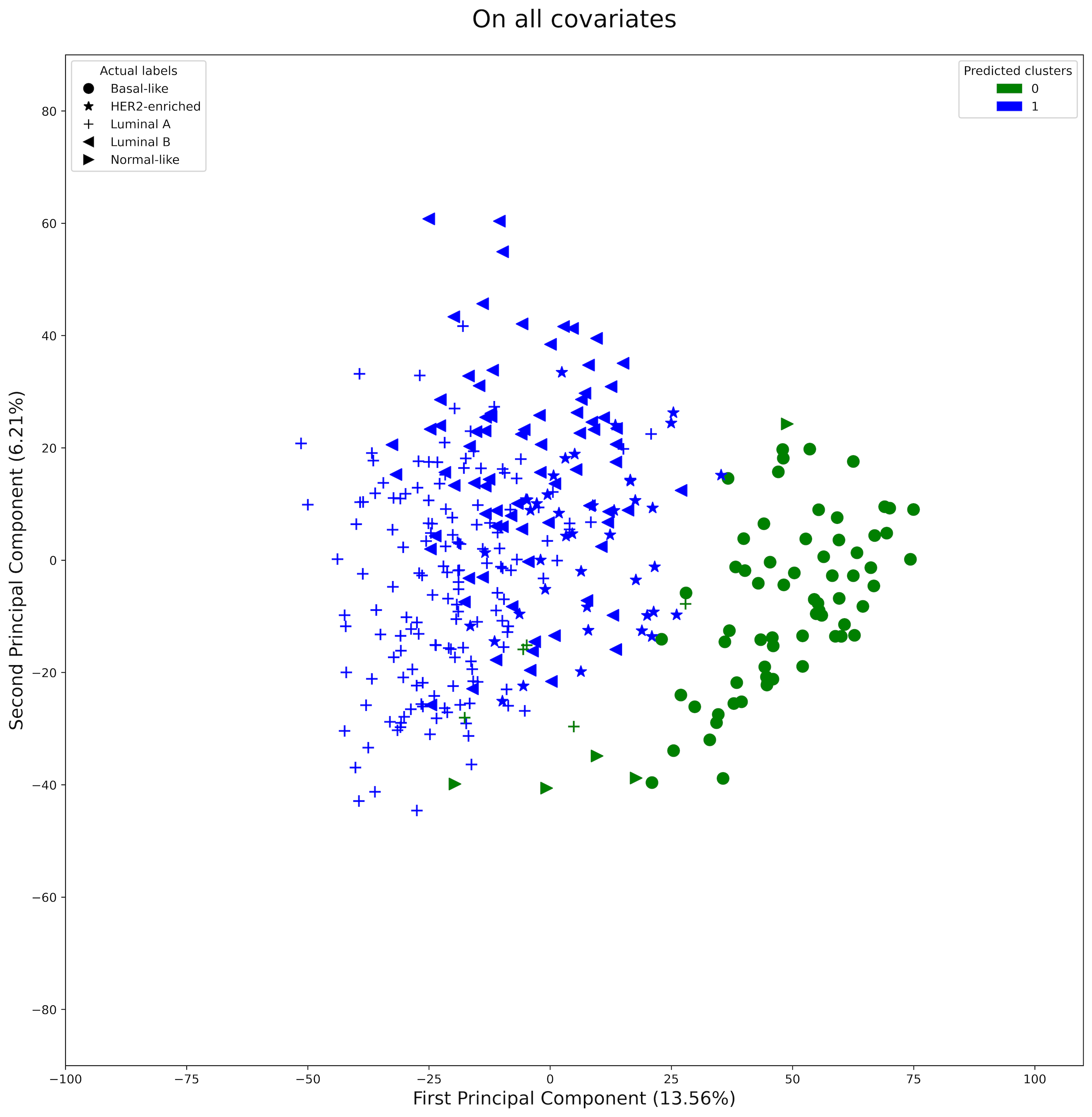}%
        }%
    \caption{\textbf{VBVarSel TCGA 2-cluster model}. This figure shows the scatter plots of VBVarSel 2-cluster model on the complete TCGA dataset, when PCA is applied to either all covariates (a), or only the selected ones (b).}
    \label{fig: scatter_cluster_alldataset_2clust all vs sel}
\end{figure}

When we set up the algorithm to prioritise a lower number of larger clusters\footnote{This can be done by either i) reducing $\alpha_0$ and increasing $b_{0j}$, which are respectively the effective prior number of observations associated with each mixture component, and the scale parameter which influences the spread of the corresponding cluster, or ii) we implement a geometric annealing schedule to strike a balance between exploration and exploitation. See Sections \ref{app-sec-methods} in the Supplementary Materials for details.}
we obtained a 2-clusters model, as shown in Figure \ref{fig: scatter_cluster_alldataset_2clust all vs sel}. In this model, a median of 6203 covariates were selected, which is similar to the 4-5 clusters model. Among the PAM50, the median rate of selection was slightly higher, approximately 82\%. Cluster 0 is significantly associated with Basal-like tumours (\textit{Fisher Test}, $p \ll 0.00001$).
In both models, variable selection produced smaller, tighter clusters and increased the separation between them.

In order to further test the robustness and accuracy of VBVarSel, we then created a semi-synthetic dataset composed by the PAM50 genes, and an increasing number of covariates selected at random from the full TCGA breast cancer dataset. The additional covariates had their rows permuted, in order to break the link between these genes and the clusters in the data, making these variables irrelevant for stratification. We then measured their probability of being selected as relevant.
The experiments showed that the variable selection machinery was able to correctly discriminate between informative covariates from non-informative ones, independently from their location in the dataset and proximity with each other, while returning a sensible stratification. The details and results of these experiments are included in Section \ref{sec:appendix-tcga} of the Supplementary Material.

\subsection*{Pan-cancer proteomic characterisation}

We complemented our TCGA data with protein expression data from The Cancer Proteome Atlas (TCPA) \citep{li2013tcpa, akbani2014pan}. TCPA provides protein expression levels across a broad spectrum of tumor and cell line samples, quantified using reverse-phase protein arrays (RPPAs) \citep{sheehan2005use}. Our TCPA dataset included 5,157 tumor samples (rows), with each sample featuring expression data for 217 proteins (columns). These proteins were pre-selected for their relevance to cancer biology and therapy \citep{akbani2014pan}. The samples are classified into 19 different cancer types.

Compared to the TCGA data, TCPA data are relatively lower in dimension, with the number of observations surpassing the number of variables. Additionally, we observe a lesser degree of correlation between covariates in TCPA. Despite the identification of 19 cancer types, there remains potential for discovering subtypes within these categories, as well as inter-cancer relationships \citep{cancer2013tcga, uhlen2017pathology, cancer2012comprehensiveBreast}, presenting a unique challenge and opportunity for our analysis. \\

\begin{figure}[htp]
    \centering
    \begin{minipage}{0.45\textwidth}
        \centering
        \includegraphics[width=\textwidth]{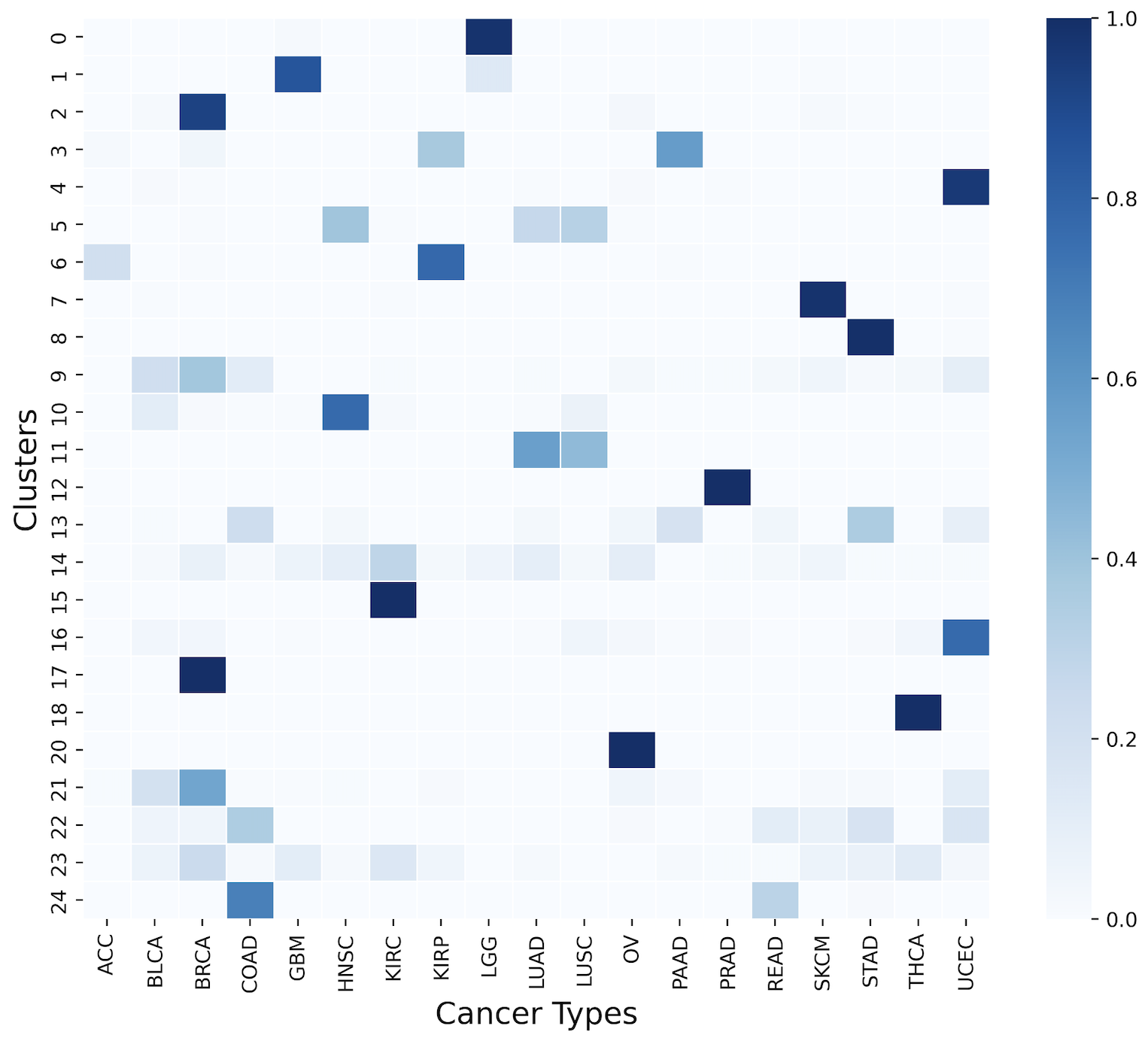}
        \subcaption{}
    \end{minipage}\hfill
    \begin{minipage}{0.55\textwidth}
        \centering
        \includegraphics[width=\textwidth]
        {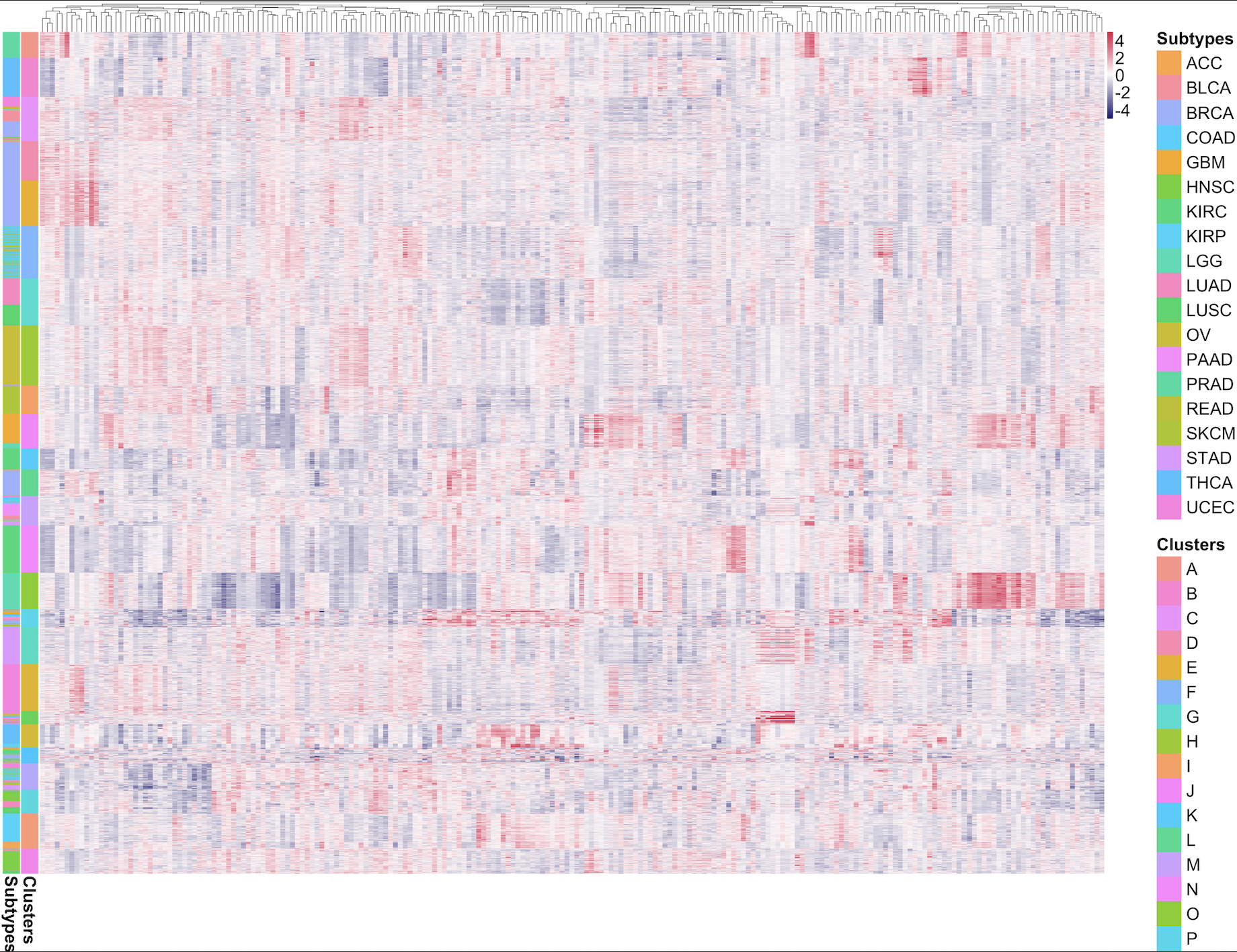}
        \subcaption{}
    \end{minipage}
    \caption{\textbf{VBVarSel clustering performance on TCPA data} (a) A heatmap illustrating the correspondence between VBVarSel clusters and tissues of origin, with clusters containing fewer than 20 observations filtered out. Each entry is calculated with respect to the tissue of origin and it indicates the percentage of observations from a given tissue assigned to each VBVarSel cluster. (b) A heatmap of the TCPA gene expression data showing the VBVarSel stratification. The annotation bars indicate the different cancer subtypes and clusters. }
    \label{fig:heatmap_tcpa_clustvssub}
\end{figure}

We applied VBVarSel to the full 5157 x 217 TCPA protein expression dataset \citep{cancer2013tcga}. The algorithm identified on average 25 clusters with more than 20 observations. Figure \ref{fig:heatmap_tcpa_clustvssub} (a) shows the correspondence between the inferred clusters and the cancer subtypes. Most cancers were strongly associated with a unique cluster. When there was some overlap, this is in agreement with other relevant literature \citep{HOADLEY2014929, akbani2014pan, CrookGattoKirk+2019}. For instance, the cancers HNSC, LUAD, and LUSC which are all aero-digestive cancers were most often grouped together. Same applies to STAD, COAD, and READ which are cancers of the digestive tract. In contrast, breast cancer (BRCA)  and endometrial cancer (UCEC) were split into subgroups \citep{akbani2014pan}. Figure \ref{fig:heatmap_tcpa_clustvssub} (b) instead shows a heatmap of the stratification and the retained genetic expressions. We observe nice agreement between clusters, cancer subtypes, and genetic expressions.

Remarkably, despite very little knowledge about the data and expected performance to inform our parameter initialisation, VBVarSel was able to converge to sensible results with ``standard'' configurations obtained using the ELBO for model selection.

As for variable selection, VBVarSel usually retained most of the variables, with a median rate of retention of 90\%. Given the profiled proteins in TCPA were already pre-selected \citep{akbani2014pan}, and we obtained sensible stratification, there is no indication that the rate is inadequate. Moreover, with different parameter initialisation, such as lower $d_0$ or $c_j$, we were able to obtain a lower retention rate but the stratification obtained was considerably worse. To better assess variable selection, we permuted the rows of varying numbers of randomly selected covariates. As expected, at least 91\% of the ``perturbed'' variables were deselected. \\

\section*{Discussion}\label{sec12}

In this work, we introduced VBVarSel, a scalable and efficient annealed variational Bayes algorithm tailored for high-dimensional mixture models, aimed at disease subtyping and biomarker discovery. Our method addresses two fundamental challenges in the current clustering literature: the curse of dimensionality and local optima trap, by incorporating annealing and variable selection within a variational inference framework. Through our simulation studies and real-world applications, we demonstrated that VBVarSel outperforms existing methodologies in terms of both runtime and accuracy, making it a robust alternative in the computational toolbox. The simulation results indicate that VBVarSel consistently outperforms traditional methods, empirically demonstrating the benefits of Variational Inference as a computationally efficient alternative to other more popular inference methods in the field. VBVarSel’s ability to infer relevant variables with minimal computational cost, while maintaining high accuracy in clustering structures, is particularly advantageous in real-world biomedical datasets where dimensionality is often high \citep{witten2010framework,review,kirk2023bayesian}. The comparison with VarSelLCM \citep{marbac2017variableVARSELLCM}, for instance, highlights that while both methods can achieve high clustering accuracy, VBVarSel is significantly faster, making it more suitable for large-scale datasets.

A key feature of our method is its ability to incorporate annealing, which helps overcome the issue of local optima by gradually smoothing the optimisation landscape. This feature is especially important in biological data, where the presence of multiple plausible clustering structures can mislead traditional variational inference methods \citep{kirk2023bayesian}. Our empirical findings supported the theoretical claims that establishing an effective balance between exploration and exploitation with a time-dependent temperature schedule would enhance inference in multi-modal posterior landscapes. Indeed, we observed increased robustness and adaptability to sub-optimal parameter initialisations, correlated data, and noise and a stabilised inference with both synthetic and real data.  

In our real-world applications to breast cancer subtyping and pan-cancer proteomics, VBVarSel exhibited strong performance, not only in terms of stratification but also in identifying relevant biomarkers. The ability to handle full, high-dimensional datasets without the need for extensive preprocessing or dimensionality reduction steps underscores the practicality of the method. Moreover, the application to the TCGA breast cancer dataset and TCPA proteomics data showcases VBVarSel’s flexibility across various cancer types, highlighting its broad applicability in precision medicine.

Despite its strengths, some limitations remain. Starting from the clustering task, while the model showed promising and sensible results on real data, pushing its stratification accuracy beyond a certain threshold was challenging. A future direction could be a semi-supervised approach, such as outcome-guided clustering. The idea is to introduce a measurable response variable, such as survival time, to guide clustering and find patterns associated with differences in outcomes. This could help improve the interpretability of stratification results in clinical settings. Another area for enhancement is our choice of the covariate selection indicator. We believe allowing both a continuous or binary indicator could offer a more nuanced understanding of each covariate's importance, particularly in datasets like TCPA, where the differences in variable salience are subtle. A flexible indicator system could provide more granular insights into how strongly each variable contributes to clustering. Finally, while the annealing process helps mitigate the local optima problem, its performance is still sensitive to the choice of temperature schedule. Careful tuning of this parameter is required, particularly in highly complex datasets. 

VBVarSel presents a significant advancement in clustering and variable selection for high-dimensional biomedical datasets, offering both scalability and accuracy. Its applications in disease subtyping and biomarker discovery, as demonstrated in this study, suggest that it can serve as a powerful tool in the era of precision medicine, where understanding the molecular underpinnings of diseases is crucial for improving patient outcomes. Future work could extend the methodology to supervised learning settings or multi-omics data integration, further broadening its applicability in biomedical research.

\backmatter

\bmhead{Supplementary information}

Supplementary Information is available for this paper.

\bmhead{Acknowledgments}
E.P. is now a doctoral student at the University of Oxford, supported by the Oxford EPSRC Centre for Doctoral Training in Health Data Science (EP/S02428X/1). 
F.P. was supported by the UK Engineering and Physical Sciences Research Council (Programme number EP/R018561/1), and is now supported by the UKRI for grant EP/Y014650/1, as part of the ERC Synergy project OCEAN. 
P.D.W.K. acknowledges core MRC (UKRI) funding through Kirk’s programme at the MRC Biostatistics Unit, as well as funding from the European Union’s Horizon 2020 Research and Innovation Programme under Grant Agreement No. 847912.

\bmhead{Code availability} 

The full VBVarSel code in Python needed to reproduce the findings of this article is available at https://github.com/MRCBSU/vbvarsel as a  package for PyPI.

\newpage

\setcounter{section}{0}
\setcounter{figure}{0}
\setcounter{table}{0}

\renewcommand{\thesection}{\arabic{section}} 
\renewcommand{\thefigure}{\arabic{figure}} 
\renewcommand{\thetable}{\arabic{table}} 

\section*{Supplementary Material}

\section{Complete VBVarSel model}\label{app-sec-methods}

 Given the data $X= \left \{ {\bf x}_n \right \}_{n=1}^N$ where $x_n$ is $J$-dimensional vector of random variables, we define the $K$-components generative mixture models as,
\begin{equation}
    p(X| \Phi,\pi) = \prod_{n=1}^N \prod_{k=1}^K \pi_k f_X({\bf x}_n | \Phi_k)
\end{equation} 
where $f_X({\bf x}_n|\Phi_{k})$ is the functional form for component $k$, parametrised by $\Phi_{k}$. In our model, we focus on GMM, which are linear combinations of Gaussian distributions, mathematically presented as follows:
\begin{equation}
    p(X| \Phi,\pi) = \prod_{n=1}^N \prod_{k=1}^K \pi_k \mathcal{N}({\bf x}_n|\boldsymbol{\mu}_{k}, \boldsymbol{\Lambda}_k^{-1})
\end{equation}
\begin{equation}
    \mathcal{N}({\bf x}_n|\mu_{k}, \boldsymbol{\Lambda}_k^{-1}) = \frac{\sqrt{|\boldsymbol{\Lambda}_k|}}{(2\pi)^{J/2}} \exp\left(-\frac{1}{2} ({\bf x}_n - \mathbf{\mu}_k)^T \boldsymbol{\Lambda}_k ({\bf x}_n - \mathbf{\mu}_k)\right)
\end{equation}
We used multivariate Gaussian distributions with parameters $\Phi_k = \left \{ \mu_k, \Lambda_k \right \}$, respectively mean vector $\mathbf{\mu}_k$ and precision matrix $\boldsymbol{\Lambda}_k$. For each observation ${\bf x}_n$ we introduce a latent variable ${\bf z}_n$, representing cluster assignment, which is a ``1-of-$K$'' binary vector of length $K$ which has precisely one non-zero element (\textit{one-hot} encoding). If $z_{nk}=1$, then ${\bf x}_n$ is associated with the $k^{th}$ component. By conditioning on the latent variable $Z$, we can decompose the joint distribution as follows:
\begin{equation}
p(X,Z,\pi,\Phi) = p(X|Z, \Phi)p(Z|\pi)p(\pi)p(\Phi)
\end{equation}
where $p(\pi)$ and $p(\Phi)$ are priors on the mixture weights and component-specific parameters (respectively). We can write down the conditional distribution of $Z$ as 
\begin{equation}
p(Z|\pi) = \prod_{n=1}^N\prod_{k=1}^K \pi_k^{z_{nk}};
\end{equation}
and similarly the conditional distribution of the observed data as
\begin{equation}\label{eq: cond dist intermediate}
p(X|Z, \Phi)= \prod_{n=1}^N\prod_{k=1}^K f_X({\bf x}_n | \Phi_k)^{z_{nk}}.
\end{equation}
In this formulation, we have implicitly assumed independence between observations and components, which allowed us to factorise in $n$ and $k$. We make another critical assumption on the independence between covariates $j$, given the component allocations $Z$, which allows us to further factorise our functional form,
\begin{equation}\label{eq: fact_j}
   f_X({\bf x}_n | \Phi_{k}) = \prod_{j=1}^J f_j({x}_{nj} | \Phi_{kj}) = \prod_{j=1}^J \mathcal{N}_j({x}_{nj}|\mu_{kj}, \tau_{kj}^{-1})
\end{equation}
Where $x_{nj}$ denotes the $j^{th}$ dimension of ${\bf x}_n$, $\Phi_{kj} = \left \{ \mu_{kj}, \tau_{kj} \right \}$ denotes the parameters associated with the $k^{th}$ mixture component, restricted to the $j^{th}$ covariate. This factorisation is equivalent to having $\Lambda_k$ as a diagonal matrix with diagonal entries $\tau_{kj}$. Our functional form is now a univariate Gaussian distribution:
\begin{equation} \label{eq:functional form}
    f_j({x}_{nj} | \Phi_{kj}) = \mathcal{N}_j({x}_{nj}|\mu_{kj}, \tau_{kj}^{-1}) = \left ( \frac{\tau_{kj}}{2\pi} \right )^{1/2}\exp\left(-\frac{1}{2} \tau_{kj} ({x}_{nj} - \mu_{kj})^2\right)
\end{equation}
We refer to our mixture model as a product of univariate - multivariate - Gaussians.

\subsection{Covariate selection model}
In our approach, model-based clustering is performed concurrently with the selection of relevant variables. In our formulation, we introduce a latent binary variable $\gamma_j \in  \left \{ 0,1\right \}$ indicating whether feature $j$ should be used to infer the clustering structure ($\gamma_j=1$) or not ($\gamma_j=0$). We name $\gamma_j$ as a covariate selection indicator. We extend Equation \eqref{eq: fact_j} as follows:
\begin{align}
f({\bf x}_n | \Phi_k, \gamma) &= \prod_{j = 1}^J f_j(x_{nj}|\Phi_{kj})^{\gamma_j}f_j(x_{nj}|\Phi_{0j})^{1 - \gamma_j} \\
 &= \prod_{j = 1}^J \mathcal{N}_j({x}_{nj}|\mu_{kj}, \tau_{kj}^{-1})^{\gamma_j}\mathcal{N}_j(x_{nj}|\mu_{0j}, \tau_{0j}^{-1})^{1 - \gamma_j}
\end{align} 
where $\Phi_{0j}$ denotes parameter estimates obtained under the null assumption that there is no clustering structure present in the $j^{th}$ covariate. We pre-compute these estimates before starting the inference procedure by Maximum Likelihood Estimate (MLE) as the mean and the precision of the data. Given the data $X= \left \{ {\bf x}_n \right \}_{n=1}^N$ where ${\bf x}_n$ is $J$-dimensional vector of random variables, for each dimension $j$ we compute:
\begin{align} \label{eq: mu0 tau0}
    \mu_{0j}^{MLE} &= \frac{1}{N} \sum_{n=1}^N x_{nj} \\
    \tau_{0j}^{MLE} &= \left ( \frac{1}{N} \sum_{n=1}^N (x_{nj} - \mu_{0j}^{MLE})^2 \right )^{-1}
\end{align}
Given the introduction of the latent variable $\gamma$, we update the decomposed joint distribution as follows:
\begin{equation}\label{eq:3.11}
p(X,Z,\pi,\Phi, \gamma) = p(X|Z, \Phi, \gamma)p(Z|\pi)p(\pi)p(\Phi)p(\gamma)
\end{equation}
where $p(\gamma)$ is the prior on the covariate selection indicators. We can write the conditional distribution of the observed data in Equation \eqref{eq: cond dist intermediate} as 
\begin{equation}\label{eq: prob dist X}
p(X|Z, \Phi, \gamma)= \prod_{n=1}^N\prod_{k=1}^K \left [ \prod_{j = 1}^J f_j(x_{nj}|\Phi_{kj})^{\gamma_j}f_j(x_{nj}|\Phi_{0j})^{1 - \gamma_j} \right ]^{z_{nk}}
\end{equation}
where the functional form $f_j$ is given in Equation \eqref{eq:functional form}.

\subsection{Prior distributions}
As we proceed to introduce the priors over the parameters $\pi$, $\Phi = \{\mu, \tau\}$, and $\gamma$, we strategically choose to work with conjugate prior distributions. 
 
We take our prior on $\pi$ to be a symmetric Dirichlet distribution with fixed concentration parameter $\alpha_0$ for each cluster, not subject to inference. 
\begin{equation}\label{eq: dirich}
p(\boldsymbol{\pi})=\operatorname{Dir}\left(\boldsymbol{\pi} \mid \boldsymbol{\alpha}_0\right)=C\left(\boldsymbol{\alpha}_0\right) \prod_{k=1}^K \pi_k^{\alpha_0-1}
\end{equation}
where $C(\alpha)$ is just the normalisation constant.  The $\alpha_0$ parameter can be interpreted as \textit{pseudocounts}, i.e.\ the effective prior number of observations associated with each mixture component. 
Given we do not want to impose a strong preliminary belief of how the proportions should be distributed, we set it to be the same for every component, meaning that all components are equally likely \textit{a priori}. The role of $\alpha_{0}$ is also crucial to automatically infer the number of clusters $K$. By setting $0<\alpha_{0}<1$, we effectively favor clustering structures in which some of the mixing coefficients are zero, i.e.\ some clusters are shrunk to zero assignments.\\

We then proceed to discuss the prior distribution on $\Phi = \{\mu, \tau\}$. Each mixture component $k$ is modeled as a product of independent univariate Gaussian distributions with parameters $\Phi_{kj} = \{\mu_{kj}, \tau_{kj}\}$. We take independent Gaussian-Gamma priors for all $\mu_{kj}, \tau_{kj}$, so that:
\begin{align} \label{eq:priorPhi}
p(\Phi_{kj}) = p(\mu_{kj}, \tau_{kj}) 	&= p(\mu_{kj}|\tau_{kj})p(\tau_{kj})\\
				&= \mathcal{N}(\mu_{kj}| m_{0kj}, (\beta_{0kj} \tau_{kj})^{-1}) \Gamma(\tau_{kj} | a_{0kj}, b_{0kj})
\end{align}
and,
\begin{align}\label{eq: gaussian-gamma}
\Gamma(\tau_{kj} | a_{0kj}, b_{0kj}) &= \frac{b_{0kj}^{a_{0kj}}}{\Gamma(a_{0kj})}\tau_{kj}^{{a_{0kj}}-1}\exp{(-b_{0kj} \tau_{kj})}
\end{align}
where $\Gamma$ is the Gamma distribution. Together these distributions constitute a Gaussian-Gamma conjugate prior distribution and their conjugacy guarantees that the posterior will take the form of a Gaussian-Gamma.  We have introduced 4 hyperparameters.  The mean parameter $m_{0kj}$ influences the center of the corresponding Gaussian distribution in the mixture, while the shrinkage parameter $\beta_{0kj}$  influences the tightness and spread of the cluster, with smaller shrinkage leading to tighter clusters. The degrees of freedom, $a_{0kj}$,  controls the shape of the Gamma distribution, the higher the degree of freedom, the more peaked (i.e.\ less dispersed) the Gamma distribution will be. Hence, $a_{0kj}$ directly influences the variability of the clusters and their overlap in the feature space. Finally, the scale parameter $b_{0kj}$ scales the Gamma distribution, the larger $b_{0kj}$, the broader the range of potential precisions, which influences the spread of the corresponding cluster. We set $\beta_0$, $a_0$ to be equal for every $j^{th}$  dimension and every $k^{th}$ cluster, while we set a $m_{0j}$ and $b_{0j}$ for every $j^{th}$  dimension. \\

For the covariate selection indicators $\gamma$, we introduce another parameter $\delta$, on which we condition to allow conjugacy. Indeed, for each $\gamma_j$, we take an independent Bernoulli conditional prior with parameter $\delta_j$, so that:
\begin{align}\label{eq: bernoulli}
p(\gamma_j|\delta_j)  = \delta_j^{\gamma_j} (1 - \delta_j)^{1 - \gamma_j},
\end{align}

The conjugate prior of a Bernoulli distribution is the Beta distribution. Hence, we take independent symmetric Beta priors for $\delta_j$, so that:   
\begin{align}\label{eq: beta}
p(\delta_j) = \mbox{Beta}(d_0).
\end{align}

The value of $\delta_j$ represents the probability of $\gamma_j = 1$. We use a symmetric Beta distribution with fixed shape parameter $d_0$, equal across every dimension $j$. The symmetry around 0.5 implies no prior preference for either $\gamma_j = 1$ or $\gamma_j = 0$. When $d_0 = 1$, the Beta distribution turns into a uniform distribution. For $d_0 < 1$, the Beta distribution is ``U-shaped'' and $\delta_j$ is more likely to take ``extreme'' values (0 or 1). For $d_0>1$ it is instead ``bell-shaped'', and middle values ($\approx 0.5$) are preferred.

\subsection{Variational distribution}
The complete joint distribution of all variables is given by:
\begin{align}\label{eq: var-model}
    p(X, Z, \pi, \mu, \tau, \gamma, \delta) = p(X| Z,  \mu, \tau, \gamma) p(Z| \pi) p(\pi) p(\mu | \tau) p(\tau) p(\gamma | \delta) p(\delta)
\end{align}
For the variational distribution, we obtain the following factorisation between parameters and latent variables:
\begin{align}\label{eq: var-dist}
    q( Z, \pi, \mu, \tau, \gamma, \delta) = q(Z)q(\pi) \prod_{j=1}^J q(\gamma_j|\delta_j)q(\delta_j)\prod_{k=1}^Kq(\mu_{kj} | \tau_{kj}) q(\tau_{kj})
\end{align}
Each factor will be updated iteratively as we minimise the KL divergence between the variational distribution and the actual posterior distribution. To derive the update equations, we utilise the foundational formula presented earlier at Equation \eqref{qUpdate}. \\ \\
\textbf{Updating $Z$}\\
Starting with the latent cluster assignments $Z$, we derive the following:
\begin{align}
\ln q^\ast(Z) = \sum_{n=1}^N \sum_{k=1}^K z_{nk} \ln \rho_{nk}+ const\label{Zupdate}
\end{align}
where we define
\begin{align}\label{rhonk}
\ln \rho_{nk} &= \mathbb{E}_\pi[\ln \pi_k] + \mathbb{E}_{\Phi, \gamma}[\ln f({\bf x}_n|\Phi_{k})]
\end{align}
Note that:
\begin{align}
\mathbb{E}_{\Phi, \gamma}[\ln f({\bf x}_n|\Phi_{k})] &= \mathbb{E}_{ \gamma}[\mathbb{E}_{\Phi}[\sum_{j=1}^J (\gamma_j \ln f_j(x_{nj}|\Phi_{kj}) + (1 - \gamma_j) \ln f_j(x_{nj}|\Phi_{0j}) )]\\
&= \sum_{j=1}^J \left(c_j \mathbb{E}_{\Phi}[\ln f_j(x_{nj}|\Phi_{kj}) ] + (1 - c_j)f_j(x_{nj}|\Phi_{0j})\right)  
\end{align}
where $c_j = \mathbb{E}_{ \gamma}[\gamma_j]$. We introduce $r_{nk}$, the responsibility of the $k^{th}$ component for the $n^{th}$ observation. 
\begin{align}
    r_{nk} = \frac{\rho_{nk}}{\sum_{k = 1}^K \rho_{nk}} = \mathbb{E}[z_{nk}]
\end{align}
Further, we make the following definition:
\begin{align}
N_k = \sum_{n=1}^N r_{nk} \label{Nk}
\end{align}
which is the expected number of observations associated with the $k^{th}$ component (note that $N_k$ need not be a whole number).\\ \\
\textbf{Updating $\pi$}\\
Next, we consider the mixing proportions $\pi$. We recognise $q^\ast(\pi)$ as an asymmetric Dirichlet distribution with parameter $\alpha = [\alpha_1, \ldots, \alpha_k]$, where 
\begin{align}
    \alpha_k = \alpha_0 + N_k
\end{align}

Recall that $N_k$ is a function of the responsibilities, $r_{nk}$, as given in Equation \eqref{Nk}.  Hence, the contribution of the covariate selection indicators on $\pi$ occurs via the responsibilities, $r_{nk}$.\\\\
\textbf{Updating $\Phi$}\\
We derive the following expression for $q^\ast(\Phi)$:
\begin{align}
\ln q^\ast(\Phi_{kj}) &= \sum_{n=1}^N {r_{nk}}c_j \ln f_j(x_{nj}|\Phi_{kj}) +  \ln p(\Phi_{kj})+ const
\end{align}
Note that we weight the contribution of the log-likelihood, $\ln f_j(x_{nj}|\Phi_{kj})$, by the factor $c_j = \mathbb{E}_\gamma[\gamma_j]$.  Hence if the $j^{th}$ covariate does not contribute to the clustering structure (i.e.\ $c_j \approx 0$), then $\Phi_{kj}$ will be dominated by the prior. Given the form of the conjugate prior on $\Phi_{kj}$ (Equation \eqref{eq:priorPhi}) and the functional form $f_j(x_{nj}|\Phi_{kj})$, we derive:
\begin{align}
    q^\ast(\Phi_{kj}) =& q^\ast(\mu_{kj} | \tau_{kj})q^\ast( \tau_{kj}) \\ 
    =& \mathcal{N}(\mu_{kj}|m_{kj}, (\beta_{kj}\tau_{kj})^{-1})\Gamma(\tau_{kj} | a_{kj}, b_{kj})
\end{align}
We introduced the following parameters ($\beta_{kj}$, $m_{kj}$, $a_{kj}$, $b_{kj}$), and statistics ($\bar{x}_{kj}$, $S_{kj}$) of the observed data, with respect to the responsibilities:

\begin{align}
    & \beta_{kj} = c_j\sum_{n=1}^N r_{nk} + \beta_{0} \\
    & m_{kj} = \frac{1}{\beta_{kj}} \left (  c_j\sum_{n=1}^N r_{nk} x_{nj} + m_{0j}\beta_{0}  \right ) \\
    & a_{kj} = \frac{1}{2} c_j\sum_{n=1}^N {r_{nk}} + a_{0} \\
    & b_{kj} = b_{0j} +\frac{1}{2} \left [   c_j N_kS_{kj} + \frac{\beta_{0}c_jN_k}{\beta_{0} +c_jN_k} \left (  \bar{x}_{kj} -  m_{0j}  \right )^2    \right] \\
    & \bar{x}_{kj} = \frac{1}{N_k}\sum_{n=1}^N {r_{nk}}x_{nj} \\
    & S_{kj} = \frac{1}{N_k} \sum_{n=1}^N {r_{nk}}(x_{nj} -\bar{x}_{kj})^2
\end{align}\\\\
\textbf{Updating $\gamma_j$} \\
For the covariate selection indicator $\gamma_j$ we unsurprisingly obtain a Bernoulli distribution, 
\begin{align}
q^\ast(\gamma_j|\delta_j) = c_j^{\gamma_j} (1 - c_j)^{1 - \gamma_j}
\end{align}
where
\begin{align}
    c_j = \frac{\eta_{1j}}{\eta_{1j} + \eta_{2j}} = \mathbb{E}_{\gamma}[\gamma_j],
\end{align}
and
\begin{align}
\ln \eta_{1j} &= \mathbb{E}_{\delta_j}[\ln (\delta_j)] + \sum_{n=1}^N \sum_{k=1}^K r_{nk} \mathbb{E}_{\Phi}[\ln f_j(x_{nj}|\Phi_{kj})]\label{nu1-main}\\
\ln \eta_{2j} &= \mathbb{E}_{\delta_j}[\ln(1 - \delta_j)] + \sum_{n=1}^N \sum_{k=1}^K r_{nk} \ln f_j(x_{nj}|\Phi_{0j})].\label{nu2-main}
\end{align} \\
\textbf{Updating $\delta_j$} \\
Next, we consider $\delta_j$, for which we obtain an asymmetric Beta distribution:
\begin{align}
q^\ast(\delta_j) = \mbox{Beta}(c_j + d_0, 1-c_j + d_0).
\end{align} \\ \\
\textbf{Evaluating $r_{nk}$ and $c_j$} \\
Having derived update equations for the variational distributions, we are left to evaluate $r_{nk}$ and $c_j$, which are the expected value of the cluster allocations and the covariate selection indicators respectively. Recall that we have:
 \begin{equation}\label{eq:exp z gamma}
     r_{nk} = \frac{\rho_{nk}}{\sum_{k = 1}^K \rho_{nk}} = \mathbb{E}[z_{nk}] \quad \textrm{and} \quad  c_j = \frac{\eta_{1j}}{\eta_{1j} + \eta_{2j}} = \mathbb{E}_{\gamma}[\gamma_j],
 \end{equation}
 where to evaluate $\rho_{nk}$, $\eta_{1j}$ and $\eta_{2j}$ as in Equations \eqref{rhonk}, \eqref{nu1-main}, and \eqref{nu2-main} respectively, we require expressions for $\mathbb{E}_\pi[\ln \pi_k]$, $\mathbb{E}_{\Phi}[\ln f_j(x_{nj}|\Phi_{kj}) ]$, $\mathbb{E}_{\delta_j}[\ln \delta_j]$, and $\mathbb{E}_{\delta_j}[\ln (1-\delta_j)]$.
We can easily write the value for $\mathbb{E}_\pi[\ln \pi_k]$ from standard properties of the Dirichlet distribution:
\begin{align}
\mathbb{E}_\pi[\ln \pi_k]  = \psi(\alpha_k) - \psi \left (\sum_{k=1}^K \alpha_k \right ), \label{eq:expPi}
\end{align}
where $\psi$ denotes the digamma function. \\ \\ We evaluate $\mathbb{E}_{\Phi}[\ln f_j(x_{nj}|\Phi_{kj})]$ as 
\begin{align}
    \mathbb{E}_{\Phi}[\ln f_j(x_{nj}|\Phi_{kj}) ] =  - \frac{1}{2} \ln 2\pi + \frac{1}{2}\mathbb{E}_{\tau_{kj}}[\ln \tau_{kj}] -\frac{1}{2}\mathbb{E}_{\mu_{kj},\tau_{kj}}[(x_{nj} - \mu_{kj})^2\tau_{kj}],
\end{align}
and
\begin{align}
    & \mathbb{E}_{\tau_{kj}}[\ln \tau_{kj}] = \psi(a_{kj}) -\ln b_{kj} \label{eq:expTau}\\ 
    & \mathbb{E}_{\mu_{kj},\tau_{kj}}[(x_{nj} - \mu_{kj})^2\tau_{kj}] = \frac{a_{kj}}{b_{kj}} (x_{nj} - m_{kj})^2 + (\beta_{kj})^{-1}. \label{eq:expPhi}
\end{align}
Finally, from standard properties of the Beta distribution we evaluate:
\begin{align}
    & \mathbb{E}_{\delta_j}[\ln \delta_j] = \psi(c_j + d_0) - \psi(2d_0+1)\label{eq:expDelta} \\ 
    & \mathbb{E}_{\delta_j}[\ln (1-\delta_j)] = \psi(1-c_j + d_0) - \psi(2d_0+1).\label{eq:expDeltaminus}
\end{align}

\subsection{Inference}

The inference process itself, which concerns the optimisation of the variational posterior distribution, can be divided into two steps, much like the EM algorithm. It begins with a \textit{variational E-step}, during which the current distributions and the current estimate of the parameters, are used to evaluate the expected values in Equations \eqref{eq:expPi}, \eqref{eq:expTau}, \eqref{eq:expPhi},  \eqref{eq:expDelta}, and \eqref{eq:expDeltaminus}. These are then used to evaluate the current estimate of the cluster assignments $\mathbb{E}[z_{nk}] = r_{nk}$, and the covariate selection indicators $\mathbb{E}[\gamma_{j}] = c_j$. Then, in the \textit{variational M-step}, $r_{nk}$ and $c_j$ are kept fixed and used to re-compute an estimate of the posterior variational distributions. The algorithm cycles through E and M steps until convergence is achieved. \\ \\
\textbf{Variational lower bound and convergence} \\
In our variational framework, we evaluate the ELBO as \footnote{To keep the notation easier, given the equation is already involved itself, we have omitted the subscripts on the expectation operator. In reality, each expectation is taken with respect to all the variables in its argument. }:
\begin{equation}
\begin{aligned}\label{eq:elbo}
\mathcal{L}= & \sum_{\mathbf{Z}} \iiint q(\mathbf{Z}, \boldsymbol{\pi}, \boldsymbol{\mu}, \boldsymbol{\tau}, \boldsymbol{\gamma}, \boldsymbol{\delta}) \ln \left\{\frac{p(\mathbf{X}, \mathbf{Z}, \boldsymbol{\pi}, \boldsymbol{\mu}, \boldsymbol{\tau}, \boldsymbol{\gamma}, \boldsymbol{\delta})}{q(\mathbf{Z}, \boldsymbol{\pi}, \boldsymbol{\mu}, \boldsymbol{\tau}, \boldsymbol{\gamma}, \boldsymbol{\delta})}\right\} \mathrm{d} \boldsymbol{\pi} \, \mathrm{d} \boldsymbol{\mu} \, \mathrm{d} \boldsymbol{\tau} \, \mathrm{d} \boldsymbol{\gamma} \, \mathrm{d} \boldsymbol{\delta} \\
= & \, \mathbb{E}[\ln p(\mathbf{X}, \mathbf{Z}, \boldsymbol{\pi}, \boldsymbol{\mu}, \boldsymbol{\tau}, \boldsymbol{\gamma}, \boldsymbol{\delta})]-\mathbb{E}[\ln q(\mathbf{Z}, \boldsymbol{\pi}, \boldsymbol{\mu}, \boldsymbol{\tau}, \boldsymbol{\gamma}, \boldsymbol{\delta})] \\
= & \, \mathbb{E}[\ln p(\mathbf{X} \mid \mathbf{Z}, \boldsymbol{\mu}, \boldsymbol{\tau}, \boldsymbol{\gamma}, \boldsymbol{\delta})]+\mathbb{E}[\ln p(\mathbf{Z} \mid \boldsymbol{\pi})]+\mathbb{E}[\ln p(\boldsymbol{\pi})]+\mathbb{E}[\ln p(\boldsymbol{\mu}, \boldsymbol{\tau})] +  \mathbb{E}[\ln p(\boldsymbol{\gamma}, \boldsymbol{\delta})]\\
& -\mathbb{E}[\ln q(\mathbf{Z})]-\mathbb{E}[\ln q(\boldsymbol{\pi})]-\mathbb{E}[\ln q(\boldsymbol{\mu}, \boldsymbol{\tau})] - \mathbb{E}[\ln q(\boldsymbol{\gamma}, \boldsymbol{\delta})].
\end{aligned}
\end{equation}

\subsection{Annealed Variational framework}
Previous derivations followed the standard variational inference procedure for a general model with latent variables. To introduce annealing in the framework, we proceed as before but start from the annealed foundational formula in Equation \eqref{eq:annealing-update}. In most cases, this only yields an additional $1/T$ factor in the parameter update. We report only the equations for the parameters updates that are directly influenced by the temperature parameter. For the latent variables $Z$ and $\gamma$, we get:

\begin{align}
\ln \rho_{nk} &= \frac{1}{T}\mathbb{E}_\pi[\ln \pi_k] + \frac{1}{T}\mathbb{E}_{\Phi, \gamma}[\ln f({\bf x}_n|\Phi_{k})]
\end{align}
\begin{align}
\ln \eta_{1j} &= \frac{1}{T}\mathbb{E}_{\delta_j}[\ln (\delta_j)] + \frac{1}{T}\sum_{n=1}^N \sum_{k=1}^K r_{nk} \mathbb{E}_{\Phi}[\ln f_j(x_{nj}|\Phi_{kj})]\\
\ln \eta_{2j} &= \frac{1}{T}\mathbb{E}_{\delta_j}[\ln(1 - \delta_j)] + \frac{1}{T}\sum_{n=1}^N \sum_{k=1}^K r_{nk} \ln f_j(x_{nj}|\Phi_{0j})]
\end{align}
\begin{align}
    &\mathbb{E}_{\delta_j}[\ln \delta_j] = \psi \left (\frac{1}{T}(c_j + d_0 + T - 1) \right) - \psi \left ( \frac{1}{T}(2d_0 + 2T - 1) \right ) \\
    &\mathbb{E}_{\delta_j}[\ln (1-\delta_j)] = \psi \left (\frac{1}{T}(T - c_j + d_0 ) \right) - \psi \left ( \frac{1}{T}(2v + 2T - 1) \right )
\end{align}
which are then used to evaluate $r_{nk}$ and $c_j$ as in Equations \eqref{eq:exp z gamma}. The annealed posterior distributions over $\pi$ and $\Phi$ are parametrised by:
\begin{align}
    & \alpha_k = \frac{1}{T} \left ( N_k + \alpha_0 + T - 1 \right ) \\
    & \beta_{kj} = \frac{1}{T} \left [ c_j\sum_{n=1}^N r_{nk} + \beta_{0j} \right ] \\
    & m_{kj} = \frac{1}{T \beta_{kj}} \left (  c_j\sum_{n=1}^N r_{nk} x_{nj} + m_{0j}\beta_{0j}  \right ) \\
    & a_{kj} = \frac{1}{T}\left [  \frac{1}{2} c_j\sum_{n=1}^N {r_{nk}} + a_{0j} + T - 1  \right] \\
    & b_{kj} = \frac{1}{T}b_{0j} +\frac{1}{2T} \left [   c_j N_kS_{kj} + \frac{\beta_{0j}c_jN_k}{\beta_{0j} +c_jN_k} \left (  \bar{x}_{kj} -  m_{0j}  \right )^2    \right]
\end{align}
And the annealed posterior distribution of $\delta$ becomes:
\begin{align}
    q^\ast(\delta_j) = \mbox{Beta}\left (\frac{1}{T}(c_j + d_0 + T - 1), \frac{1}{T}(T - c_j + d_0 ) \right )
\end{align}

The annealed variational lower-bound is as in Equation \eqref{eq:elbo} but with an additional $T$ factor in front of the negative terms and is also indirectly affected by the updated annealed parameters. Importantly, when $T = 1$, we retrieve the standard (non-annealed) variational inference.

\section{Summary of prior specifications}

We provide a summary of prior specifications which we found worked well for the given datasets. 

\begin{table}[htp!]
\centering
\begin{tabular}{lrrrrrrrrrr}
\toprule
Experiment & $K$ & $\alpha_0$ & $m_{0j}$ & $\beta_{0j}$ & $a_{0j}$ & $b_{0j}$ & $d_0$ & $c_j$ \\
\midrule
Synthetic & [3,10] & [0.1, 1] & mean($X_j$) & $10^{-3}$ & 3. & [0.1, 1] & 0.9 & [0.5, 1]  \\
TCGA  & [5, 8] & [0.01, 0.1] & mean($X_j$) & $10^{-3}$ & 3. & [0.1, 1] & [1, 5] & 1  \\
A-TCGA & [5, 7] & $1/K$ & mean($X_j$) & $10^{-3}$ & [3, 10] & [0.1, 1] & [0.9, 5] & 1  \\
TCPA & [25, 40] & $10^{-3}$ & mean($X_j$) & $10^{-3}$ & 3. & 0.1 & 0.5 & [0.8, 1]  \\
\bottomrule
\end{tabular}
\captionsetup{justification=centering}
\caption{\label{prior init} \textbf{Parameter initialisations.} For some parameters we provide fixed values, for others a range of values that worked well. We omit $z_{nk}$ and $\delta_j$ as we always sample them from the corresponding distributions.  }
\end{table}
\textbf{Legend}:
\begin{itemize}
    \item $K$: maximum number of clusters in the overfitted mixture.
    \item $\alpha_0$: concentration of the Dirichlet prior on the mixture weights $\pi$ (Eq. \eqref{eq: dirich})
    \item $m_{0j}$ and $\beta_{0j}$: mean and shrinkage of the Gaussian conditional prior on the components mean $\mu_{kj}$ (Eq. \eqref{eq:priorPhi})
    \item $a_{0j}$ and $b_{0j}$: degrees of freedom and scale of the Gamma prior on the components precision $\tau_{kj}$ (Eq. \eqref{eq:priorPhi})
    \item $d_0$: shape of the Beta prior on the covariate selection probabilities $\delta_j$ (Eq. \eqref{eq: beta})
    \item $c_j$: covariate selection indicator
    \item $z_{nk}$: cluster assignment
\end{itemize}

\section{Simulation Study: additional experiments}

In this section we provide additional experiments on the simulated dataset to further evaluate VBVarSel's clustering ad feature selection performance when using annealing.

\subsection{Prior specifications for simulation studies}

For VBVarSel, our experiments showed that the model is quite robust to the initialisation of $\beta_{0}$, $m_{0j}$, $a_0$ and $K$. On the contrary, the model performance was influenced by the concentration parameter $\alpha_0$, the shape parameter $d_0$, and most significantly the scale $b_{0j}$. Starting with $\alpha_0$, this parameter strongly affected the ability to ``empty'' extra clusters. Nonetheless, any value $<0.5$ consistently allowed convergence to the true $K$ in this simulated environment. As for $d_0$, values $<0.5$ led to higher deselection rate, and the opposite is true for $d_0>5$; in between the performance was stable on perfect selection. Most importantly, VBVarSel requires very careful tuning of $b_{0j}$. Even slight deviations from optimal would significantly and detrimentally impact the quality of the stratification.

\subsection{Evaluating benefits of annealing}

We demonstrate the benefits of annealing in overcoming common challenges faced in real-world data scenarios, such as correlated data, sub-optimal parameter initialisation, and noise. We explore different temperature schedules. We begin with $T=2$, $3$ or $4$, which either remain constant throughout inference or follow the geometric or harmonic schedule. For time-varying schedules, we set 5 or 10 maximum annealed iterations, given we normally converge in less than 15 iterations. We report the performance of the annealing approaches that allowed more significant advantages, and also the non-annealed model ($T=1$) for reference.

We simulate \citep{CrookGattoKirk+2019} synthetic data with $n=100$ observations and 200 variables, of which 20 (10\%) are relevant. In order to show the benefits of annealing, we make this simulated data more realistic by first introducing correlation. Instead of using identity variance-covariance matrices to generate relevant variables, we introduce different degrees of correlation, i.e.\ off-diagonal non-zero entries. Table \ref{annealing simCrook-corr-rand-all} reports the performance of VBVarSel when randomly sampling the correlation for each cluster and each covariate between 0 and 0.5. Table \ref{annealing simCrook-corr-fix} reports the performance of VBVarSel with fixed and equal covariance across all dimensions in all components, and Table \ref{annealing simCrook-corr-rand} with randomly sampled correlation for each cluster, but fixed across all covariates. All experiments were run with \textit{optimal} parameter initialisations and results are averaged across 10 independent runs.

\begin{table}[htp!]
\centering
\begin{tabular}{lrrrrr}
\toprule
Temperature & Relevant & Irrelevant & ARI \\
\midrule
 $T=1$  & 1 [1, 1] & 0.99 [0.99, 0.99] & 0.48 [0.41, 0.54] \\
$T= 2 \text{G}$  & 1 [1, 1]& 1 [1, 1]& 0.69 [0.69, 0.71] \\
\hline
 $T=2$  & 1 [1, 1] & 1 [1, 1] & 0.59 [0.40, 0.71] \\

\bottomrule
\end{tabular}
\captionsetup{justification=centering, font=footnotesize}
\caption{\label{annealing simCrook-corr-rand-all} Annealed VBVarSel performance on synthetic data modified to include randomly sampled covariances across all clusters and relevant covariates. G: Geometric schedule and the initial temperature.}
\end{table}

Across all varying degrees of introduced covariance, we observe a general improvement with annealing. This enhancement manifests in several aspects, whether it is an improved stratification or variable selection accuracy, or increased stability across experiments. This is even more pronounced when we amplify the randomness and variability in the correlation structure (Table \ref{annealing simCrook-corr-rand-all}). Indeed, as the stochasticity in the correlation structure increases, we observe that implementing an effective exploration-exploitation balance with a geometric schedule becomes more beneficial.  Notably, the geometric schedule we used is relatively straightforward, thus demonstrating that annealing does not require intensive fine-tuning efforts to show its benefits in a simulated environment.

\begin{table}[htp!]
\centering
\begin{tabular}{llrrrr}
\toprule
Covariance & Temperature & Relevant & Irrelevant & ARI \\
\midrule
\multirow{2}{*}{0.1}& $T=1$  & 1 [1, 1] & 1 [0.99, 1] & 0.97 [0.97, 0.97] \\
                    & $T= 2 \text{H}$  & 1 [1, 1] & 1 [1, 1] & 1 [0.97, 1] \\
\hline
\multirow{2}{*}{0.5} & $T=1$  & 1 [1, 1] & 1 [0.99, 1] & 0.68 [0.50, 0.71] \\
                     & $T= 3 \text{G}$  & 1 [1, 1] & 1 [1, 1] & 0.76 [0.76, 1] \\
\hline
\hline
0.1 & $T=2$  & 1 [1, 1]  & 1 [1, 1] & 1 [1, 1]  \\
0.5 & $T=2$  & 1 [1, 1] & 1 [1, 1] & 0.70 [0.70, 0.73] \\
\hline
\end{tabular}
\caption{\label{annealing simCrook-corr-fix} Annealed VBVarSel performance on synthetic data modified to include fixed covariance. G: Geometric, H: Harmonic schedule and the initial temperature is given.}
\end{table}

\begin{table}[htp!]
\centering
\begin{tabular}{lrrrrr}
\toprule
Temperature & Relevant & Irrelevant & ARI \\
\midrule
 $T=1$  & 1 [1, 1] & 1 [0.99, 1] & 0.65 [0.65, 0.70] \\
$T= 2 \text{G}$  & 1 [1, 1] & 1 [1, 1] & 0.74 [0.74, 1] \\
\hline
$T=2$  & 1 [1, 1] & 1 [1, 1] & 0.71 [0.67, 1] \\

\bottomrule
\end{tabular}
\captionsetup{justification=centering, font=footnotesize}
\caption{\label{annealing simCrook-corr-rand}Annealed VBVarSel performance on synthetic data modified to include randomly sampled covariances for each cluster. G: Geometric schedule and the initial temperature is given.}
\end{table}

Table \ref{annealing simCrook}  show the performance of VBVarSel with sub-optimal parameter initialisations. We refer to \textit{optimal} parameter initialisation as the one used in previous simulations, reported in Table \ref{prior init}, experiment \textit{synthetic}. As for the \textit{sub-optimal} initialisation, we vary the scale $b_{0j}$ since it is the parameter to which VBVarSel is more sensitive. We randomly choose a value for $b_{0j}$ between $0.01$ and $1$ in each of the 10 randomisations of the data we ran, and we report the median scores with upper and lower quartiles.

Table \ref{annealing simCrook-noise} show the performance of VBVarSel with added Gaussian noise.
Starting from the original \citep{CrookGattoKirk+2019} synthetic dataset, we add Gaussian noise with zero mean and a varying standard deviation (noise level). Even though in real-world scenarios the noise might not always follow a Gaussian distribution, it is a sensible approximation, providing a good balance between simplicity and realism.

We observe how little changes in the initialisation affect the performance of the non-annealed VBVarSel but introducing annealing generally allowed the optimiser to better explore the posterior space and ultimately reach the global optimum. Furthermore, across all varying Gaussian noise levels, although the VBVarSel algorithm is already reasonably robust to noise, introducing even a straightforward temperature schedule yields improved performance and stability, without increasing the computational complexity of the model.

\begin{table}[ht]
\centering
\begin{tabular}{llrrrr}
\toprule
Initialisation & Temperature & Relevant & Irrelevant & ARI \\
\midrule
\multirow{3}{*}{Optimal}& $T=1$  & 1 [1, 1] & 1 [1, 1] & 1 [1, 1] \\
                        & $T= 3 \text{G}$ & 1 [1, 1] & 1 [1, 1] & 1 [1, 1] \\
                        & $T= 2 \text{H}$ & 1 [1, 1] & 1 [1, 1] & 1 [1, 1] \\
\hline
\multirow{3}{*}{Sub-optimal}& $T=1$  & 1 [1, 1] & 0.98 [0.97, 0.99] & 0.84 [0.75, 0.88] \\
                            & $T= 3 \text{G}$ & 1 [1, 1] & 1 [1, 1] & 1 [0.70, 1] \\
                            & $T= 2 \text{H}$ & 1 [1, 1] & 1 [1, 1] & 1 [0.94, 1] \\
\hline
\hline
Optimal & $T=2$  & 1 [1, 1] & 1 [1, 1] & 1 [1, 1] \\
Sub-optimal & $T=2$ & 1 [1, 1] & 1 [1, 1] & 1 [0.84, 1] \\

\bottomrule
\end{tabular}
\caption{\label{annealing simCrook} Annealed VBVarSel performance on \citep{CrookGattoKirk+2019} synthetic data using \textit{optimal} and \textit{sub-optimal} parameter initialisations. G: Geometric, H: Harmonic schedule and the initial temperature is given. }
\end{table}

\begin{table}[htp!]
\centering
\begin{tabular}{llrrrr}
\toprule
Noise Level & Temperature & Relevant & Irrelevant & ARI \\
\midrule
\multirow{3}{*}{0.1} & $T=1$  & 1 [1, 1] & 0.98 [0.98, 1] & 0.89 [0.86, 0.95] \\
                     & $T= 3 \text{G}$  & 1 [1, 1] & 1 [1, 1] & 1 [0.93, 1] \\
                     & $T= 3 \text{H}$  & 1 [1, 1] & 1 [1, 1] & 1 [1, 1] \\
\hline
\multirow{3}{*}{0.5}& $T=1$  & 1 [1, 1] & 0.98 [0.97, 0.98] & 0.90 [0.65, 0.92] \\
                    & $T= 2 \text{G}$  & 1 [1, 1] & 1 [1, 1] & 1 [0.77, 1] \\
                    & $T= 2 \text{H}$  & 1 [1, 1] & 1 [1, 1] & 1 [1, 1] \\
\hline
\hline
0.1 & $T=2$  &  1 [1, 1] &  1 [1, 1]& 1 [1, 1] \\
0.5 & $T=4$  &  1 [1, 1] &  1 [1, 1] &  1 [0.70, 1] \\
\bottomrule
\end{tabular}
\caption{\label{annealing simCrook-noise} Annealed VBVarSel performance on synthetic data modified to include Gaussian noise. We averaged across 10 independent runs. G: Geometric, H: Harmonic schedule and the initial temperature is given.}
\end{table}

\subsection{Model misspecification}

We generated synthetic data in two scenarios: one where we added Student's t-distributed noise to our simulated data, and another where the relevant variables data directly followed a Student's t multivariate distribution. 

OPTION 1: Gaussian multivariate dist but with student t noise. Clusters generated as in Crook, student t noise added with DoF  [2, 3, 3] ( no normalisation)

\begin{table}[htp!]
\centering
\begin{tabular}{lrrrrrr}
\toprule
Method & $n$ & \% & Time (seconds) & Relevant & Irrelevant & ARI \\
\midrule
VBVarSel & 100 & 10 & 3.2 [2.5, 5.2] & 1 [1, 1] & 0.99 [0.98, 1] & 0.60 [0.58, 0.68] \\
 &  & 25 & 4.2 [1.9, 5.6] & 1 [1, 1] & 1 [1, 1] & 0.56 [0.49, 0.65] \\
 &  & 50 & 2.7 [1.6, 4.6] & 1 [1, 1] & 0.99 [0.99, 1] & 0.46 [0.40, 0.55] \\
\hline 
VBVarSel & 1000 & 10 & 10.5 [10.2, 13.6] & 1 [1, 1] & 1 [1, 1] & 0.69 [0.67, 0.73] \\
 &  & 25 & 9.8 [9.1, 11.2] & 1 [1, 1] & 1 [1, 1] & 0.78 [0.72, 0.88]  \\
 &  & 50 & 8.4 [8.0, 12.6] & 1 [1, 1] & 1 [1, 1] & 0.69 [0.66, 0.78]  \\
\bottomrule
\end{tabular}
\caption{\label{tab:misspecification 1} }
\end{table} 

OPTION 2: Student t multivariate. similar configs as crook simulation df : 3 ( normalisation)

\begin{table}[htp!]
\centering
\begin{tabular}{lrrrrrr}
\toprule
Method & $n$ & \% & Time (seconds) & Relevant & Irrelevant & ARI \\
\midrule
 VBVarSel & 100 & 25 & 4.6 [2.6, 5.2] & 0.82 [0.72, 0.9] & 1 [0.99, 1] & 0.68 [0.62, 0.78] \\
 &  & 50 & 4.6 [2.4, 5.3] & 0.96 [0.85, 0.99] & 1 [1, 1]  & 0.74 [0.71, 0.80] \\
\hline 
VBVarSel & 1000 & 25 & 11.2 [9.1, 13.7] & 0.92 [0.69, 0.98] & 1 [1, 1] & 0.59 [0.57, 0.63]  \\
 &  & 50 & 10.1 [9.3, 15.6] & 0.88 [0.86, 0.93] & 1 [1, 1] & 0.58 [0.56, 0.68]  \\
\bottomrule
\end{tabular}
\caption{\label{tab:misspecification 2} }
\end{table}

\section{TCGA Data: additional experiments} \label{sec:appendix-tcga}

In this section we provide additional experiments on the breast cancer transcriptomic data from The Cancer Genome Atlas (TCGA) \citep{cancer2013tcga} to further evaluate VBVarSel's clustering ad feature selection performance.

\subsection{Unsupervised model-based clustering on PAM50 genes}

We assessed the clustering capabilities of VBVarSel, while temporarily neglecting variable selection. We extract only the PAM50 genes from the full dataset, which should constitute only relevant information, we fix the covariate selection indicators at 1, and disable inference on those.

When using a scale $b_{0j} = 1$, we obtained 5 clusters that reasonably resemble the breast cancer subtypes (ARI $\approx 0.54$). Cluster A is associated with Luminal A samples, while Cluster C is mostly associated with Luminal B samples. Cluster B contains only HER2-enriched samples, but it gathers those that are most ``distant'' in feature space from Luminal B. Cluster D perfectly represents Basal-like samples, and Cluster E seems to have identified the Normal-like samples despite very few occurrences. The overlap between clusters in feature space, which is due to an existing similarity in some genetic expressions, presents the most significant challenge to our model's accuracy. However, the results obtained are aligned with established literature \citep{CrookGattoKirk+2019, cancer2012comprehensiveBreast}.

When instead initialising lower $b_{0j}$ and $\alpha_0$, the algorithm converged to a 2-clusters model, one containing only basal-like samples, and the other grouping together the remaining types. This stratification was indeed maximising the ELBO.

\subsection{Simultaneous stratification and biomarker selection on semi-synthetic breast cancer data}
Having established that our model's clustering performance aligns with the current literature and expected outcomes, we proceed to enable variable selection. We maintain the PAM50 genes in our dataset, but also progressively add covariates randomly selected from the full dataset and permute the rows of these. The aim is to disrupt or ``break'' the existing clustering information in the permuted genes, making them irrelevant for stratification, and observe if the model discards them. In fact, while PAM50 genes are backed by research as relevant markers, it's crucial to understand that not all other genes are inherently irrelevant. However, by artificially rendering additional genes irrelevant in this semi-synthetic environment, we aim to rigorously assess our algorithm's ability to discern truly informative variables from noise.

The dataset used always includes the PAM50 genes, to which we add progressively $p=$ 50, 100, and 500 genes to simulate scenarios where the number of covariates is either lower, comparable or larger than the number of observations. 

Table \ref{pam50+random} shows the resulting averaged performance of the VBVarSel algorithm for varying numbers of  $p$ additional randomly sampled and permuted genes. The first row reports the results obtained with PAM50 genes only.  We report the time, in seconds, the number of retained PAM50 genes, together with the total number of relevant and irrelevant variables, and the ARI between the inferred stratification and the \textit{ground-truth} cancer subtype of each observation. We present the median scores with the upper and lower quartiles across 10 independent runs on different data randomisation. 

\begin{table}[htp!]
\centering
\begin{tabular}{rrrrrr}
\toprule
p  & Time (seconds) & Relevant &  Irrelevant & ARI \\
\midrule
0 & 2.10 [1.50, 3.15] & 44 [43, 45] &  6 [5,  7] & 0.39 [0.38, 0.49] \\
50 & 3.27 [2.82, 4.09] & 41 [39, 43]  &  59 [57, 61] &  0.41 [0.40, 0.45]  \\
100 & 5.20 [4.88, 6.17] & 49 [47, 50]   & 101 [100, 103] & 0.41 [0.38, 0.43]\\
500 & 23 [17.9, 40.9] & 62 [58, 65]  & 488 [485, 492] & 0.38 [0.34, 0.44] \\
\bottomrule
\end{tabular}
\caption{\label{pam50+random} VBVarSel performance on varying subsets of TCGA data. }
\end{table}

Statistically, the retention of PAM50 genes is significantly better than random across all experiments (\textit{Fisher Test}, $p <  0.00001$). The stratification quality remains constant, demonstrating that VBVarSel scales well with increasing number of covariates and the performance stays approximately constant amidst noisy variables. The algorithm also scales very well in terms of runtime.

\subsection{Benchmarking on pre-processed TCGA expression dataset}

To allow comparison with existing literature, we pre-process the complete TCGA dataset  as in \citep{lock2013preprocess} and \citep{CrookGattoKirk+2019}. We keep 645 genes for each of the 348 tumour samples, of which 14 are from the PAM50 subset.  Initialising lower $\alpha_0$ and $b_{0j}$,  VBVarSel converges to two clusters, and 318 variables are selected to discriminate between the two groups, which includes all the 14 PAM50 genes (\textit{Fisher Test}, $p < 0.00005$). These results are comparable to what is reported in \citep{CrookGattoKirk+2019}, although VBVarSel tends to select more variables overall, and they are also in agreement with stratifications and selection rates obtained in previous experiments. Indeed, we again observe smaller, tighter, and more separable clusters when focusing on the retained variables.

\newpage



\bibliography{sn-bibliography}

\end{document}